\documentclass[journal]{IEEEtran}
\DeclareFixedFont{\auacc}{OT1}{phv}{b}{it}{18}   
\DeclareFixedFont{\newauacc}{OT1}{ptm}{b}{rm}{12}   
\usepackage{hyperref}
\usepackage{cite}
\usepackage{graphicx}
\usepackage{subfigure}
\usepackage{url}
\usepackage{theorem}
\usepackage{framed}
\usepackage{arydshln}
\usepackage[reqno]{amsmath}
\usepackage{amsfonts}
\usepackage{amssymb}
\usepackage{color}
\usepackage{paralist}
\usepackage{verbatim}
\DeclareMathSymbol{\R}{\mathord}{AMSb}{"52}
\DeclareMathSymbol{\C}{\mathord}{AMSb}{"43}
\DeclareMathSymbol{\Z}{\mathord}{AMSb}{"5A}
\DeclareMathSymbol{\N}{\mathord}{AMSb}{"4E}
\DeclareMathSymbol{\K}{\mathord}{AMSb}{"4B}
\DeclareMathSymbol{\M}{\mathord}{AMSb}{"4D}
\DeclareMathSymbol{\Q}{\mathord}{AMSb}{"51}
\DeclareMathSymbol{\Lset}{\mathord}{AMSb}{"4C}

\theorembodyfont{\rmfamily}
\theoremheaderfont{\bfseries\slshape}
\newtheorem{theorem}{Theorem}

\theoremheaderfont{\rmfamily}

\newtheorem{lem}{Lemma}

\theorembodyfont{\upshape}

\newtheorem{remark}{Remark}

\newcommand\defeq{\mathrel{:=}}

\newcommand\ie{{\it i.e.}}
\newcommand\eg{{\it e.g.}}
\newcommand\etal{{\it et al.}}
\newcommand{\ud}{\mathrm{d}}

\definecolor{webmag}{rgb}{0.5,0,0.5}
\newcommand{\nop}[1]{}

\newcommand{\rmH}{\textrm{H}}
\newcommand{\rmL}{\textrm{L}}
\newcommand{\rmR}{\textrm{R}}
\newcommand{\rmC}{\textrm{C}}
\newcommand{\rmX}{\textrm{X}}

\title{On the Asymptotic Validity of the Decoupling Assumption for Analyzing 802.11 MAC Protocol}
\author{\authorblockN{Jeong-woo Cho, Jean-Yves Le Boudec,~\IEEEmembership{Fellow,~IEEE,} and Yuming Jiang}
\thanks{This work was supported in part by ``Centre for Quantifiable Quality of Service in Communication Systems, Centre of Excellence'' appointed by The Research Council of Norway, and funded by The Research Council, NTNU and UNINETT. A part of this work was done when J. Cho was with EPFL, Switzerland.}
\thanks{J. Cho is with the School of Information and Communication Technology at KTH Royal Institute of Technology, Stockholm, Sweden (email: jwcho@kth.se).}
\thanks{J.-Y. Le Boudec is with \'Ecole Polytechnique F\'ed\'erale de Lausanne (EPFL), CH-1015 Lausanne, Switzerland (email: jean-yves.leboudec@epfl.ch).}
\thanks{Y. Jiang is with the Department of Telematics, Norwegian University of Science and Technology (NTNU), N-7491, Trondheim, Norway. (email: jiang@item.ntnu.no).}
}

\makeatletter
\let\@copyrightspace\relax
\makeatother

\begin{document}
\maketitle

\newcommand{\expectation}{\textsf{E}}
\newcommand{\probability}{\textsf{P}}
\newcommand{\pdf}{\textsf{f}}
\newcommand{\slow}{\ell}
\newcommand{\nextline}{\mbox{}\\}
\newcounter{tempcounter}
  \setlength{\belowcaptionskip}{0pt}

\begin{abstract}
Performance evaluation of the 802.11 MAC protocol is classically based on the decoupling assumption, which hypothesizes that the backoff processes at different nodes are independent. This decoupling assumption results from mean field convergence and is generally true in transient regime in the asymptotic sense (when the number of wireless nodes tends to infinity), but, contrary to widespread belief, may not necessarily hold in stationary regime. The issue is often related with the existence and uniqueness of a solution to a fixed point equation; however, it was also recently shown that this condition is not sufficient; in contrast, a sufficient condition is a global stability property of the associated ordinary differential equation. In this paper, we give a simple condition that establishes the asymptotic validity of the decoupling assumption for the homogeneous case. We also discuss the heterogeneous and the differentiated service cases and formulate a new ordinary differential equation. We show that the uniqueness of a solution to the associated fixed point equation is not sufficient; we exhibit one case where the fixed point equation has a unique solution but the decoupling assumption is not valid in the asymptotic sense in stationary regime.
\end{abstract}


\section{Introduction}\label{sec:intro}

\IEEEPARstart{T}{he} Wireless LAN standard is evolving towards higher and higher aggregate throughput. The increased maximum bit rate of 802.11n, $600$Mbps, along with its easy deployability, suggests the potential use of an 802.11n access point as an wireless router transacting a huge amount of data of many nodes. In this work, we focus on the performance evaluation of 802.11 under the many-node regime as the population size (the number of wireless nodes) $N$ tends to infinity.

Most existing work on performance evaluation of the 802.11 MAC protocol \cite{refBianchi,refGaretto,refKumar07,refRamaiyanMultistability} relies on the ``decoupling assumption'' which was first adopted in the seminal work by Bianchi \cite{refBianchi}. Though having been defined in various ways, it essentially assumes that all the nodes in the same network experience the same {\it time-invariant} collision probability, with the direct consequence that the backoff processes are {\it independent}\footnotemark. This assumption is unavoidable primarily because the stationary distribution of the original Markov chain cannot be explicitly written due to the irreversibility of the chain \cite{refKumar07} even for small number of backoff stages, \ie, $3$ and $4$, unless the population of the network is very small. A similar point was stressed by P. R. Kumar in an interview with Science Watch Newsletter \cite{refKumarNewsletter}:

\vskip 3pt \noindent { \rmfamily\itshape ''A good analogy is in thermodynamics. Instead of trying to study the behavior of just three or four molecules and how they move around, you study the behavior of billions and trillions of molecules. \ldots~Similarly, we want to see what you can say about wireless networks in the aggregate.''} \vskip 3pt

\noindent which suggests an analogy of the intractable small-scale problems in different areas. If we {\it liken} each wireless node to a particle in a physical system, which condition would suffice for every particle being absolutely {\it decoupled} from the rest?

\footnotetext{The meaning of ``to decouple'' in the literature as well as in our work is an abuse of terminology, in the sense that it has implied not only `to decouple nodes' (independence) but also `to have a time-invariant collision probability'.}

Once we assume that the decoupling assumption holds, the analysis of the 802.11 MAC protocol leads to a {\it fixed point equation} (FPE) \cite{refKumar07},
also called Bianchi's formula. Kumar \etal \cite{refKumar07} revisited the FPE and made several remarkable observations, advancing the state of the art to more systematic models and paving the way for more comprehensive understanding of 802.11. Above all, one of the key findings of \cite{refKumar07}, already adopted in the field \cite{refRamaiyanMultistability,refKwak}, is that the full interference model, also called the single-cell model \cite{refKumar07} and the main focus of our work, leads to the {\it backoff synchrony property} \cite{refProutierePushing} which implies the backoff process can be completely separated and analyzed solely through the FPE technique. 

This decoupling assumption can be formally justified as a consequence of convergence to mean field and of Sznitman's result \cite{sznitman1991topics}; it can thus only be asymptotically true as the population $N$ goes to infinity. However, it is recently pointed out by Bena\"im and Le Boudec \cite[Section 8.2]{refJYMF} that Sznitman's result and convergence to mean field imply the asymptotic validity of the decoupling assumption only in the transient regime, \ie, over a finite horizon, and given some initial conditions. In stationary regime, there may be no decoupling assumption even in the limit of large population size $N$. This may happen for example when the ordinary differential equation (ODE) that defines the mean field limit has a limit cycle. In such a case, nodes are asymptotically independent only conditional to the state of the fluid limit. In contrast, if the ODE satisfies a strong global stability property, namely, it has a unique stationary point to which all trajectories converge, then the decoupling assumption is also valid in stationary regime \cite{refJYMF}. For the case of the 802.11 MAC protocol, the stationary points of the ODE are the solutions of the FPE mentioned above. However, existence and uniqueness of a solution to the FPE does not guarantee that all trajectories of the ODE converge to the unique fixed point; in \cite{refJYMF}, there is a simple example of mean field limit where the FPE has a unique solution but trajectories of the ODE do not converge, in general, to this unique fixed point. Therefore, though the decoupling assumptions that underly Bianchi's formula is plausible and intuitive, the question of its validity can be asked.
%
The main purpose of this paper is to provide an answer to the following question.

\smallskip
\noindent { \mdseries\slshape ``Under which conditions is the decoupling assumption for the model of the 802.11 MAC asymptotically valid?''}
\smallskip

To put it another way, we ask whether the FPE method and Bianchi's assumption are valid. To this end, we use mean field theoretic results \cite{refBordenaveMF,refJYMF,refSharmaScaledMarkov} which state that, as $N$ tends to infinity, a scaled version of the original Markov chain model of the backoff process in 802.11 MAC protocol converges to a nonlinear {\it ordinary differential equation} (ODE) so that the asymptotic validity of the decoupling assumption and thus of the FPE boils down to the stability of this ODE. Denoting by $p_k$ the attempt probability of each wireless node at each time-slot in backoff stage $k \in \{ 0,1,\cdots,K\}$, we assume in what follows that our mean field models are derived when $K$ is finite and fixed while the number of nodes $N$ goes to infinity.

 In connection with the mean field models, it is worth while to clarify why the relevant works \cite{refBordenaveMF,refJYMF,refSharmaScaledMarkov} have used a specific intensity scaling regime, under which the activity of each node in backoff stage $k$ is scaled as follows:
 \begin{align}\label{eq:intscaling}
 p_k \defeq \epsilon(N) \cdot q_k \quad \mbox{($q_k$ is a constant.)}
 \end{align}
 where $q_k$ is called the {\it scaled}~attempt rate throughout this paper. It is natural to assume that $\epsilon(N)$ is {\it vanishing}, \ie, $\lim_{N\to\infty} \epsilon(N) = 0$. Otherwise, the collision probability between wireless nodes converges to one as $N$ goes to infinity. More importantly, we have to use an appropriate form of intensity scaling $\epsilon(N)$ in order to avoid exceptional cases. For example, if $\epsilon(N)$ decreases faster than $1/N$ (\eg, $\epsilon(N)=1/N^2$), it can be easily seen that the collision probability vanishes as $N$ tends to infinity, irrespective of whichever backoff stage each node belongs to (we refer to Section \ref{sec:intensity} for a formal argument). In other words, each node is completely decoupled from the rest. On the other hand, if $\epsilon(N)$ decreases slower than $1/N$ (\eg, $\epsilon(N)=1/\sqrt{N}$), the collision probability becomes one as $N$ goes to infinity. That is to say, $\epsilon(N)=1/N$ is the {\bf only} intensity scaling regime (up to a constant factor) that deserves to be analyzed.

 Under the intensity scaling regime $\epsilon(N)=1/N$, Bordenave \etal~in \cite[Theorem 5.4]{refBordenaveMF} studied the {\bf homogeneous} case (all nodes have the same per-stage backoff probabilities) for the case when the number of backoff stages is infinite. They found the following sufficient condition for global stability of the ODE, hence for the asymptotic validity of the decoupling assumption:
 \setcounter{tempcounter}{\value{equation}}
\setcounter{equation}{0}
\renewcommand{\theequation}{BMP}
 \begin{equation}
 q_0 < \ln 2 \; \; \mbox{ and }
 q_{k+1} = q_{k}/2, \;\;\forall
k \geq 0  \label{eq-bord}
 \end{equation}
 where $q_k$ is the scaled attempt rate in \eqref{eq:intscaling} for a node in backoff stage $k$.
In this paper, we focus on the case where the total number of backoff stages $K+1$ is finite, as this is true in practice and in Bianchi's formula. Sharma \etal \cite{refSharmaScaledMarkov} obtained a result for $K=1$ and mentioned the difficulty to go beyond. A comprehensive summary of the literature and the outstanding questions raised therein has been recently made by Duffy \cite{refDuffySummary}. 
\renewcommand{\theequation}{\arabic{equation}}
\setcounter{equation}{\value{tempcounter}}

%

We find that not only (i) the monotonicity (\eqref{eq:mono} in Section \ref{sec:justification}) but also (ii) the mild intensity of scaled attempt rates (\eqref{eq:mint} in Section \ref{sec:justification}) imply the uniqueness of a solution to the FPE, which is naturally a necessary condition for stability. Moreover, we prove that the latter \eqref{eq:mint} guarantees the global stability of the ODE. Thus the condition that the attempt rate is {\it upper-bounded} by the reciprocal of the population, namely $q_k \leq 1$ for all $k$, suffices for the validity of the decoupling assumption. Moreover, for the familiar parameter setting $q_k = q_0 /m^k$ where $m\geq 1$, the condition \eqref{eq:mint} suffices for maximizing the aggregate throughput of the network, hence it is a practical condition.

In order to offer various services to higher priority users with additional performance requirements, 802.11e standard introduced the enhanced distributed channel access (EDCA) functionality that has three mechanisms to differentiate the per-class settings of (i) channel holding time, (ii) contention window (CW), and (iii) idle time after each transmission, where the first one has no effect on the backoff processes. Since the second one, CW differentiation, necessarily implies there are two or more classes, we call the corresponding system {\bf heterogeneous}. The third one, called AIFS differentiation, imposes an additional complexity on the Markov chain analysis because whether the users of a class may attempt transmission at each time-slot depends on the type of the current time-slot, which again depends on the activity of the users in the previous time-slot. This mutual interaction of the two evolutions substantially complicates the analysis. As of now, there is {\itshape no} ODE in the literature which models AIFS differentiation using an appropriate formalism.

To tackle this problem, it is of importance to observe that the stage evolution of all nodes (or stage density) is much {\it slower} than the evolution of the type of time-slots under the AIFS differentiation. Thus the former can be taken to be constant by the latter. An application of mean field theoretic result \cite[Theorems 1 \& 2]{refJYMF}, formalized based on the same observation, yields an {\it extended ODE} model of the backoff processes in EDCA-enabled 802.11 networks. We also {\bf formulate} an {\it extended FPE} on the basis of this ODE, which is satisfied by the equilibrium points of the ODE. It is remarkable that this FPE coincides with that proposed in \cite[Section VI]{refKumar07}.

The versatility of the ODE model is demonstrated by investigating some selected counterexamples. In the first example, we consider a homogeneous system where all nodes use the same parameters and show that the system is {\it bistable} in that the backoff process, after whirling closely around an equilibrium for a very long time, suddenly jumps into another equilibrium, and {\it vice versa}. The FPE model is only capable of identifying three equilibrium points as its solutions, whereas the ODE model is further capable of classifying the two of them into locally stable points and the other into unstable point, accurately reflecting the multistability. The trajectories of the ODE constitute a separatrix which divides the initial condition space into two regions. We also consider a heterogenous system where the set of nodes are divided into two classes. A delicate determination of the parameters renders the system {\it oscillatory} such that all trajectories converge to a stable limit cycle formed around an unstable unique equilibrium point where the limit cycle is as determined by the extended ODE. This example also serves
as an illustration of the fact that there may be a unique solution to the fixed point equation whereas the decoupling assumption does {\bf\itshape not} hold in the asymptotic sense.
We also stress that the stability condition established in this work for the first time has been tantalizing other researchers as well, \eg, \cite[Appendix B]{refSharmaScaledMarkov}.

The rest of the paper is organized as follows. In Section \ref{sec:justification}, we present a brief overview of recent advances in mean field theory and introduce the associated ordinary differential equation thereof. In Section \ref{sec:asympdecoupling}
, we prove a global stability condition of the ODE, which is in turn shown to be capable of optimizing the throughput. In Section \ref{sec:mfdiff}, we elaborate on another complexities arising from EDCA and derive its corresponding ODE model. Some counterexamples in Section \ref{sec:examples} illustrate the utility of the ODE models. Concluding remarks and an outstanding problem are given in Section \ref{sec:conclusion}.

\section{Mean Field Technique Revisited}\label{sec:justification}


To begin with, it should be noted that our analytical model of 802.11 MAC protocol is different from the original one. Thus we first briefly describe the original operation of 802.11 MAC in Section \ref{sec:dcf} and explore the differences between our model and the real 802.11 MAC protocol in Section \ref{sec:difference}.

 If the duration of per-stage backoff is taken to be geometric (which is uniform in the standard), the backoff process in 802.11 is governed by a few rules: (i) every node in backoff stage $k$ attempts transmission with probability $p_k$ for every time-slot; (ii) if it succeeds, $k$ changes to $0$; (iii) otherwise, $k$ changes to $(k+1)$ \verb#mod# $(K+1)$ where $K$ is the index of the highest backoff stage. Markov chain models, which have been widely used in describing complex systems including 802.11, however, very often lead to excessive complications as discussed in Section \ref{sec:intro}. In this section, we present a surrogate tool for the analysis, {\it mean field theory}. It is noteworthy that the rules used in 802.11, \ie, (i)--(iii), closely resemble the mean field equations laid out below.

\subsection{Basic Operation of DCF Mode}\label{sec:dcf}

 Time is slotted. Since our analysis is mainly focused on the backoff procedure of 802.11 distributed coordination function (DCF), we call the standardized time interval in the backoff procedure of the 802.11 standard {\it time-slot}\footnote{This is equivalent to {\it slot} in the work by Kumar \etal \cite{refKumar07} (\eg, $20\mu s$ in IEEE 802.11b).} for brevity. The durations of frames, packets, and inter-frame spaces used in the other procedures are generally different from that of a time-slot.

Each node follows the randomized access procedure of 802.11 DCF. To begin with, each node generates a {\it backoff value} if it has a data packet to send. Since the backoff procedure of each node is controlled by {\it inter-frame spaces} that fill in spaces between frames and packets, we introduce them here to help to understand the basic operation of DCF mode. Two types of inter-frame spaces are used in 802.11 DCF, namely, Short Inter-Frame Space (SIFS) and Distributed Inter-Frame Space (DIFS). Each node freezes (stops) the countdown procedure {\it as soon as} the medium becomes busy. On the other hand, only when the medium is idle for the duration of a Distributed Inter-Frame Space (DIFS), a node may unfreeze (start) its countdown procedure of the backoff and decrements the backoff by one per every time-slot. If the backoff reaches zero, the sender transmits an RTS (ready to send) frame, followed by a CTS (clear to send) from the receiver, a data packet from the sender and an ACK packet from the receiver if RTS/CTS mechanism is switched on. Note that SIFS is smaller than DIFS so that no node is allowed to interrupt a sequence of frames and packets which are spaced out SIFS apart.

    There exist $K+1$ backoff stages whose indices belong to the set $ \{0,1,\cdots,K\}$ where $K>0$. If a node has not attempted transmission for a data packet yet, the node is supposed to be in the initial backoff stage where the backoff value is drawn uniformly from $\{$0, 1, $\cdots$, 2${b_0}-$1$\}$ (or $\{$1, 2, $\cdots$, 2${b_0} \}$). Here $2 b_0$ is the contention window that serves as the initial value of a backoff countdown. If two or more wireless nodes finish their countdowns at the same time-slot, there occurs a collision between RTS frames if the RTS/CTS mechanism is switched on, otherwise two or more data packets collide with each other. If there is a collision, each node who participated in the collision multiplies its contention window by the multiplicative factor $m=2$. In other words, each node changes its backoff stage index $k$ to $k+1$ and adopts a new contention window $2b_{k+1} =2 m^{k+1} b_0 $. If $k+1$ is greater than the index of the highest backoff stage number, $K$, the node steps back into the initial backoff stage and the contention window is set to $2b_0$.

Let $L$ and $L_c$ denote the average duration of a successful packet transmission and the fixed duration of a collision, expressed in terms of backoff time-slot. Note here that the length of data packets can be arbitrary random values. Also the fixed overhead for each successful transmission is denoted by $L_o$. Note that $L$, $L_c$ and $L_o$ do not need to be integer numbers but can be arbitrary positive {\it real} numbers. In 802.11 DCF, if the RTS/CTS mechanism is used, $L$ represents the time to transmit an RTS frame, a CTS frame, a data packet, and an ACK packet plus inter-frame spaces, \ie, SIFS and DIFS, where $L_o$ is $L$ minus the time to transmit a data packet. The duration $L_c$, much smaller than $L$, is the time to transmit an RTS frame plus one DIFS.

\subsection{Differences between Our Model and 802.11 DCF Mode}\label{sec:difference}

Our analysis is made tractable by a number of differences between our model used in this paper and the original operation of 802.11 DCF mode. First of all, we take the duration of per-stage backoff to be geometric as we did at the beginning of Section \ref{sec:justification}. Secondly, the parameter set is fixed for each version of the standard whereas each parameter in our model may be an arbitrary number. For example, in the IEEE 802.11b standard, $m=2$, $K=6$ ($7$ attempts per packet), and $2b_0 = 32$ are used.

We also make a few assumptions for tractable analysis.
\begin{compactitem}
 \item {\bf Single-cell assumption}: Most importantly, this work focuses on the performance of {\it single-cell} 802.11 networks in which all 802.11-compliant nodes are within such a distance from each other that a node can hear whatever the other nodes transmit. Since all nodes freeze their backoff countdown during channel activity, the total time spent in backoff countdowns up to any time is the same for all nodes. Therefore, it is {\it sufficient} to analyze the backoff process in order to investigate the performance of single-cell networks. This technique has been adopted in many works including \cite{refKumar07,refRamaiyanMultistability,refBordenaveMF,refJYMF}.
 \item {\bf Greedy-node assumption}: Secondly, we only consider the case of greedy wireless nodes that persistently contend for the wireless medium.
\item {\bf Error-free channel assumption}: Lastly, we assume that the wireless channels are error-free so that failed transmissions are caused only by collisions between RTS frames (for the case of RTS/CTS) or data packets.
\end{compactitem}

\subsection{Bianchi's Formula}\label{sec:revisited}

In performance analysis of 802.11, Bianchi's formula and its many variants are probably the most known \cite{refBianchi,refCarvalho,refGaretto,refKumar07,refKwak,refMedepalli,refRamaiyanMultistability,refSakuraiDelay}. Assuming that there are $N$ nodes, Bianchi's formula can be written compactly in a more general {\it fixed point equation} (FPE) form:
\begin{align}\label{eq:NFPE1}
\bar p  & =  \frac{\sum_{k=0}^{K} \gamma^k}{\sum_{k=0}^{K} \frac{\gamma^k}{p_k}}, \\
\gamma  & = 1 - (1-\bar p)^{N-1} \label{eq:NFPE2}
\end{align}
where ${\bar p} $ and $\gamma$ respectively designate the average attempt probability and collision probability of every node at each time-slot. The attempt probability in backoff stage $k$ is denoted by $p_k$ and defined as the inverse of the mean contention window, \ie, $p_k = 1/(b_k -1/2)$. Note that, as long as the backoff stage $k=0$ follows backoff stage $k=K$ for any attempts, the statistics like $\bar p$ and $\gamma$ are not affected by whether attempts in the highest backoff stage $K$ are successful or not.

 The FPE model has been used as a {\it de facto} principal tool for the analysis of the 802.11 MAC Protocol. The weak point of the FPE model is that it cannot be concluded entirely from the form of FPE whether its solution (even if it is unique) might be a good first-order approximation of $\bar p$ and $\gamma$. 

Exactly under which condition the FPE holds is recently being investigated with rigorous mathematical arguments \cite{refSharmaScaledMarkov,refBordenaveMF,refJYMF}, called {\it mean field independence}. This fundamental approach was originally developed in the two works by Bordenave \etal \cite{refBordenaveOld} and Sharma \etal \cite{refSharmaInfocom} where the {\it first} mean field analyses of the 802.11 MAC protocol were performed\footnote{The conference versions of the two works were {\it submitted} at roughly the same times.}. In the rest of the paper, we will refer to their journal versions \cite{refBordenaveMF,refSharmaScaledMarkov}. Remarkably, Bordenave \etal \cite{refBordenaveMF} provided a broader mean field framework which extends to multiple-cell networks ({\it cf}. single-cell assumption in Section II-B) and supports the notion of `resource'. The particle interaction model proposed in \cite{refJYMF} overcomes some limitations and broadens applicability of the model proposed in \cite{refBordenaveMF}. The three works \cite{refSharmaScaledMarkov,refJYMF,refBordenaveMF} have found that, as the number of particles goes to infinity, \ie, $N\to \infty$, the stage distribution of every node evolves according to a set of $K+1$ dimensional nonlinear ordinary {\it differential} equations (ODE) under an appropriate scaling of time. 

\subsection{The Mean Field ODE model}\label{sec:valodemodel}

Let us dive into the details of the mean field interaction model for 802.11 used in \cite{refSharmaScaledMarkov,refBordenaveMF,refJYMF} and how the Markov chain of the model converges to the associated ordinary differential equation.

\smallskip
\noindent\underline{\bf Model description}: In our version of 802.11 DCF mode under the assumption made in Section \ref{sec:difference}, there are $N$ wireless nodes evolving in a finite state space $\{ 0, \cdots, K \}$ at discrete time-slots $t \in \{0,1,\cdots\}$. Denoting by $X_n (t) \in \{ 0, \cdots, K \}$ the backoff stage (the state) of node $n \in \{1,\cdots, N\}$ at time-slot $t$, we collect the observations $X_n (t)$, for all $n \in \{1, \cdots,N\}$ and compute the relative frequencies, which is called the occupancy measure (or empirical measure). Formally, the occupancy measure in backoff stage $k$ at (discrete) time-slot $t$ is defined as
\begin{align}\label{eq:occ}
\Phi_k (t) \defeq \frac{1}{N} \sum_{n=1}^{N} 1_{\{X_n (t) = k\} }
\end{align}
where $1_{\{\cdot \}}$ is the indicator function.  Let $\boldsymbol{A}^{\mathbf{T}}$ be the transpose of a matrix $\boldsymbol{A}$. It can be readily observed that the occupancy measure vector ${\boldsymbol \Phi}( t) \defeq ( \Phi_0(t) ~ \cdots ~ \Phi_K (t))^{\mathbf{T}}$ possesses the {\it Markovian} property
 because all nodes in the same backoff stage are exchangeable under the greedy-node assumption in Section \ref{sec:difference}. Thus the system can be described by $K$-dimensional vector ${\boldsymbol \Phi}( t)$ rather than $N$-dimensional vector ${\boldsymbol X}( t) \defeq ( X_1(t) ~ \cdots ~ X_N (t))^{\mathbf{T}}$ though the nodes are not distinguishable any more.

This Markov chain (discrete-time Markov Process) is in fact analogous to the special continuous-time Markov process, \ie, {\it density dependent population process}, that was used in the seminal work by Kurtz \cite[Chapter 11]{refKurtzDensity}. Basically, the three works, \cite{refSharmaScaledMarkov,refJYMF,refBordenaveMF}, are nontrivial extensions of the result in \cite{refKurtzDensity} to Markov chain version by means of the following scaling technique.

\smallskip
\noindent\underline{\bf Key scalings}: Unlike the density dependent population process in \cite{refKurtz}, our Markov chain in \eqref{eq:occ} cannot converge to an ODE as $N \to \infty$ because a Markov chain evolves at discrete time-slots $t \in \{0,1,\cdots\}$. The ODE is derived by means of the following two key scalings.
\begin{compactitem}
 \item {\bf Intensity scaling} is to {\it slow down} the evolution of each node by a factor of $\epsilon(N)$, such that each node in backoff stage $k$ attempts transmission with probability $p_k = \epsilon(N) \cdot q_k $.
  \item {\bf Time acceleration} is to {\it accelerate} the evolution of time-slots by $1/\epsilon(N)$, such that a variable at $t$ before this operation is translated into another variable at $t \cdot \epsilon(N)$.
\end{compactitem}

The main purpose of using the intensity scaling $\epsilon(N)$ is to make sure that the {\bf intensity}, defined as the number of state (backoff stage) transitions {\it per node} per time-slot, vanishes, \ie, converges to $0$ as $N\to \infty$. In our context, the intensity is $p_k = \epsilon(N) \cdot q_k $, and thus we require
$$ \lim_{N\to\infty} \epsilon(N) = 0 .$$
In Section \ref{sec:intensity}, the implications of intensity scaling will be explored in detail in conjunction with the collision probability and its {\it physical} meaning.

Since the intensity in the above vanishes, the number of state transitions of {\it all nodes} per time-slot is order of $N \cdot \epsilon(N)$ which is dominated by $N$. That is, the expected change of $ \Phi_k( t)$ over two consecutive time-slots is order of $\epsilon(N)$ which tends to zero as $N\to\infty$. However, if we accelerate the evolution of time-slots by $1/\epsilon(N)$, the change of $ \Phi_k( t)$ becomes order of one, and thus the time-slots get closer, hence the time continuity.  The limit variables which we obtain by applying the time acceleration and the limit operation $N \to \infty$ are dubbed {\it mean field limits (MFL)} in this paper.

To avoid notational confusion, we use capital Greek letters, $\Phi_k(\cdot)$ (or ${\boldsymbol \Phi}( \cdot)$) and $\Gamma$, to denote the original variables and lower-case letters, $\phi_k$ (or ${\boldsymbol \phi}( \cdot)$) and $\gamma$ to denote their MFLs.

\smallskip
\noindent\underline{\bf The ODE}: The scaled version of the Markov chain converges to an ODE system as $N\to \infty$.
 It is shown
in \cite{refBordenaveMF} that, as $N$ tends to infinity, ${\boldsymbol \Phi}( t/\epsilon(N)) = ( \Phi_0(t/\epsilon(N)) ~ \cdots ~ \Phi_K (t/\epsilon(N)))^{\mathbf{T}}$ converges in
probability to ${\boldsymbol\phi}(t) \defeq ( \phi_0 (t) ~ \cdots ~ \phi_K (t))^{\mathbf{T}}$ which is the solution of the ODE:
\begin{align}
\frac{\ud \phi_0}{\ud t} (t) & = {\bar q}(t)  \left( 1 - \gamma(t) \right)  - {q_0} \phi_0(t) + \underbrace{ q_K \phi_{K} (t) \gamma(t)}_{\textrm{inflow from }K}  \label{eq:homo0}
\end{align}
which is the differential equation with respect to $\phi_0(t)$ and
\begin{align}
\frac{\ud \phi_k}{\ud t} (t) & = q_{k-1} \phi_{k-1} (t) \gamma(t) - q_k \phi_k(t),~  \label{eq:ODE}
\end{align}
which is the differential equation for $k \in \{ 1,  \cdots,K\}$.
Note that we denote by
\begin{align}\label{eq:averageq}
\bar q (t) \defeq \sum_{k=0}^{K} q_k \phi_k (t)
\end{align}
the MFL of the average attempt rate and $\gamma(t)$ is the MFL of the collision probability to be defined very soon. It is important to note that the above system is {\it degenerate}\footnotemark, because we also have a {\it manifold} relation $\phi_0 (t) \equiv 1 - \sum_{k=1}^K \phi_k(t)$, which can be plugged into \eqref{eq:ODE} to eliminate \eqref{eq:homo0}, whereupon we only need to consider the $K$-dimensional system \eqref{eq:ODE}. 
We will use the reduced version \eqref{eq:ODE} throughout this work to simplify the exposition. This system \eqref{eq:ODE} will be called {\bf homogeneous} because all nodes adopt the same parameter set $q_k$ and $K$.

\footnotetext{A degenerate system has a {\it singular} Jacobian matrix which means that its linearization cannot determine the local stability of the system.}

 The differential equation \eqref{eq:ODE} can be intuitively understood. For example, the first term and second term on the right-hand side in \eqref{eq:ODE} are respectively the inflow caused by collisions in the $(k-1)$th backoff stage and the outflow caused by any attempts in the $k$th backoff stage. Note that the underbraced term in \eqref{eq:homo0} was not considered in \cite{refBordenaveMF}, and exists only in networks with finite backoff stages.

\smallskip
\noindent\underline{\bf Collision probability}: The full derivation of the ODE is omitted due to the space limit and a detailed one can be found in the works by Sharma \etal \cite[Section III]{refSharmaScaledMarkov} and Bordenave \etal \cite[Section 5]{refBordenaveMF}. However, we believe that the readers can grasp the main idea by looking into the derivation of the MFL of collision probability in the following.

Pick a backoff stage $k' \in \{0,\cdots,K\}$. For any node in backoff stage $k'$, the collision probability of the node at time-slot $t$ is given by
\begin{align}\label{eq:gammaorig}
 \Gamma(t,k') \! \defeq \! 1 \! - \! \left( \!1\! - \!\epsilon(\!N\!) q_{k'} \!\right)^{-1} \prod_{k=0}^{K} \left(1 - \epsilon(\!N\!) \cdot q_k \right)^{N  \Phi_k( t)}.
\end{align}
This is the probability that at least one other node attempts transmission at time-slot $t$. Here we can see that the term $\epsilon(N) q_{k'}$ vanishes as $N\to\infty$. 
Thus we can define the MFL of collision probability as follows:
\begin{align}\nonumber
 \gamma(t) \defeq \lim_{N\to \infty} \Gamma(t/\epsilon(N),k').
\end{align}
We assume the following special intensity scaling regime throughout the rest part of this paper:
$$\epsilon(N)=\frac{1}{N}.$$
It follows from the definition of exponential function
\begin{align}\label{eq:exp}
\lim_{N \to \infty} (1-x/N)^N = \exp(-x)
\end{align}
 and the definition of $\bar q(t)$ in \eqref{eq:averageq} that
\begin{align}
\gamma(t) = 1 - {\rm e}^{- \sum_{k=0}^{K} q_k \phi_k (t)} = 1 - {\rm e}^{- {\bar q(t)}} \label{eq:collmf}
 \end{align}
Also, remark that $\Gamma(t,k)$ depends on backoff stage $k$, whereas its MFL $\gamma(t)$ is common to all nodes.

\subsection{The Intensity Scaling Regime}\label{sec:intensity}

Here we expatiate upon our discussion on the intensity scaling $\epsilon(N)$ in Section \ref{sec:intro}. Note that the expression \eqref{eq:collmf} holds if and only if $ \epsilon(N) \in \Theta(1/N)$. When $ \epsilon(N) \notin \Theta(1/N)$, \eg, $ \epsilon(N) = 1/N^2 $ or $ \epsilon(N) = 1/\sqrt{N} $, we can see from the forms of \eqref{eq:gammaorig} and \eqref{eq:exp} that $\gamma(t)$ becomes either zero or one because $K$ is a finite constant. Summing up, if we consider $\epsilon(N) \notin \Theta(1/N)$, the decoupling assumption is asymptotically valid, which is in line with what intuition tells us.

In connection with the above discussion, the intensity scaling technique can be construed as an essential property that must be imposed upon all practical systems where particles (or nodes) {\it share} a common resource of fixed capacity \cite{refBordenaveMF}. If $\epsilon(N)$ decreases faster than $1/N$, \eg, $\epsilon(N) = 1/N^2$, the common resource is not used at all as population tends to infinity, hence no collision. On the other hand, if $\epsilon(N)$ decreases slower than $1/N$, \eg, $\epsilon(N) = 1/\sqrt{N}$, the common resource is utterly squandered in {\it attempting} transmission as $N$ tends to infinity, ending up with collisions all the time. Therefore, the physical meaning of $ \epsilon(N) = 1/N $ is crystal clear.

\subsection{Equilibrium Points}\label{sec:eqpoint}

Equating the right-hand sides of \eqref{eq:homo0} and \eqref{eq:ODE} to zero yields the following equilibrium points: \begin{equation} \nonumber
\phi_k = \frac{q_{0}}{q_k} \gamma^k \phi_{0} , ~\mbox{and} ~\phi_0 = \frac{\bar q }{q_0 \sum_{k =0}^K \gamma^k }
\end{equation}
whereupon the backoff stage distribution of every node at the equilibrium can be computed as:
\begin{equation}\nonumber
 \phi_k = \frac{ \gamma^k }{ q_k \sum_{j=0}^{K} \frac{\gamma^j}{q_j} } .
\end{equation}
By plugging the manifold relation $\sum_{k=0}^K \phi_k(t) \equiv 1$ into the above, we can get the following fixed point equation in the stationary regime:
\begin{align}\label{eq:FPE1}
\bar q  & =  \frac{\sum_{k=0}^{K} \gamma^k}{\sum_{k=0}^{K} \frac{\gamma^k}{q_k}}, \\
\gamma  & = 1 - {\rm e}^{- {\bar q}}. \label{eq:FPE2}
\end{align}
Note that, under the intensity scaling regime, \ie, $p_k = q_k/N$ and $\bar p = \bar q /N$, \eqref{eq:NFPE2} becomes
\begin{align}
 \lim_{N\to\infty} 1 - (1-\bar p)^{N-1}  =\lim_{N\to\infty} 1 - \left(1-\frac{\bar q}{N}\right)^{N-1} = 1 - {\rm e}^{- {\bar q}} \nonumber
\end{align}
which is identical to \eqref{eq:FPE2}. That is, we do not need to distinguish between \eqref{eq:FPE2} and \eqref{eq:NFPE2} under the intensity scaling regime.

\section{Asymptotic Validation of Decoupling}\label{sec:asympdecoupling}

The theoretical limit of mean field analysis represented by \eqref{eq:ODE} needs to be clearly understood. The nonlinear ODE model only implies that {\it any} node will be in backoff stage $k$ with the common probability $\phi_k(t)$ under the asymptotic regime. The component ratio ${\boldsymbol\phi}(t) = ( \phi_0 (t) ~ \cdots ~ \phi_K (t))^{\mathbf{T}}$, in general a time-varying solution of \eqref{eq:ODE}, is not guaranteed to be constant.
Bordenave \etal \cite[Theorem 5.4]{refBordenaveMF} studied its global stability of the asymptotic case when $K = \infty$, but the more practical case for finite $K$ remains to be proved. The need of a proof for finite $K$ is stressed in \cite[pp.833]{refJYMF} due to its practical implication. In line with this, by appealing to a Lyapunov function, Sharma \etal~proved this for the case $K=1$ where there are only two backoff stages \cite[Lemma 3]{refSharmaScaledMarkov}.

\subsection{Main Results}\label{sec:mainresults}

Before presenting the result for finite $K$ in Theorem~\ref{th:meanfield}, we describe {\it two} different sufficient conditions for the uniqueness of the equilibrium. 
To simplify the exposition, we first define two conditions:
\setcounter{tempcounter}{\value{equation}}
\setcounter{equation}{0}
\renewcommand{\theequation}{MONO}
\begin{align}
q_k \mbox{ is nonincreasing in } k \mbox{.}
 \label{eq:mono}
\end{align}\vspace{-5mm}
\renewcommand{\theequation}{UNIQ}
\begin{align}
\mbox{\eqref{eq:FPE1}-\eqref{eq:FPE2} has a unique solution.} \label{eq:uniq}
\end{align}
The following lemma holds as long as the right-hand side of \eqref{eq:FPE2} is increasing in $\bar q$. That is, the lemma does not fully exploit the exponential form of \eqref{eq:FPE2}. It is remarkable that Lemma \ref{lem:fixedpoint} was originally established by Kumar \etal \cite[Theorem 5.1]{refKumar07}. We give a {\it simpler} alternative proof in Appendix \ref{sec:proofpropfixed1} based on the method of mathematical induction.

\begin{lem}[Monotonicity Implies Uniqueness]\label{lem:fixedpoint} \nextline
\eqref{eq:mono} implies \eqref{eq:uniq}.
\end{lem}
\renewcommand{\theequation}{\arabic{equation}}
\setcounter{equation}{\value{tempcounter}}

To present the second sufficient condition for the uniqueness of the equilibrium, we define another condition:
\begin{align}
 \bar q (t) \leq 1,\quad \forall t \geq0 .
 \label{eq:gint}
\end{align}
As we are interested in {\it global} stability, we need to show that the solutions of \eqref{eq:ODE} with {\it any} initial condition converge to the unique equilibrium. Recall that $\bar q(t) =\sum_{k=0}^K q_k \phi_k(t)$, from the form of which it is clear that \eqref{eq:gint} holds for any initial condition $\boldsymbol\phi(0)$ {\it if and only if}
\setcounter{tempcounter}{\value{equation}}
\setcounter{equation}{0}
\renewcommand{\theequation}{MINT}
\begin{align}
 q_k \leq 1, ~\forall k.
 \label{eq:mint}
\end{align}
We call this condition \eqref{eq:mint} which is an acronym for `Mild INTensity'. In a sense, we can interpret the intensity scaling regime $p_k = q_k / N$ as a way of weakening the node activity. From this point of view, \eqref{eq:mint} implies we impose an additional constraint upon the node activity.

Interestingly, the above upper bound on the scaled attempt rates also implies \eqref{eq:uniq}. This {\it intermediate} result is presented here to shorten the proof of Theorem \ref{th:meanfield}, which is the final form of the result.
\begin{lem}[Mild Intensity Implies Uniqueness]\label{lem:fixedpoint2} \nextline
\eqref{eq:mint} implies \eqref{eq:uniq}.
\end{lem}
\renewcommand{\theequation}{\arabic{equation}}
\setcounter{equation}{\value{tempcounter}}
\begin{proof}
Putting $q_{\rm max} \defeq \max_{k\in \{0,\cdots,K \}} q_k$, it is clear that \eqref{eq:mint} is equivalent to $q_{\rm max} \leq 1$. First, we have
\begin{align}
\bar q =  \frac{\sum_{k=0}^K \gamma^k}{\sum_{k=0}^K \frac{\gamma^k}{q_k}} \leq \frac{\sum_{k=0}^K \gamma^k}{\sum_{k=0}^K \frac{\gamma^k}{q_{\rm max}}} = q_{\rm max} \leq 1.
\end{align}
Multiplying the both sides of \eqref{eq:FPE1} by ${\rm e}^{-\bar q}$ yields:
\begin{align}
\bar q {\rm e}^{-\bar q} & =  \frac{\sum_{k=0}^K \gamma^k}{\sum_{k=0}^K \frac{\gamma^k}{q_k}} \cdot  {\rm e}^{-\bar q}  = \frac{1}{\sum_{k=0}^K \frac{\gamma^k}{q_k}} \cdot \frac{ \sum_{k=0}^K \gamma^k}{\sum_{k=0}^{\infty} \gamma^k} \label{eq:multexp}
\end{align}
where the last equality follows from \eqref{eq:FPE2}, \ie, ${\rm e}^{-\bar q} = 1-\gamma = 1/\sum_{k=0}^{\infty} \gamma^k$.
The second factor of the last equation of \eqref{eq:multexp} can be rearranged as
\begin{align}
\textstyle
 \sum_{k=0}^K \gamma^k  (1-\gamma) & \textstyle = (1+\cdots+\gamma^K) - ( \gamma + \cdots+ \gamma^{K+1}) \nonumber \\ & \textstyle = 1 - \gamma^{K+1}  = 1 - (1 - {\rm e}^{-\bar q})^{K+1} \nonumber
\end{align}
which is a decreasing function of $\bar q$. As the first factor of the last equation of \eqref{eq:multexp} is also a decreasing function of $\bar q$, \eqref{eq:multexp} is decreasing in $\bar q$. On the other hand, $\bar q {\rm e}^{-\bar q}$ is increasing in $\bar q \in [0, 1]$ and the range of $\bar q {\rm e}^{-\bar q}$ is $[0, {\rm e}^{-1}]$. Since \eqref{eq:multexp} decreases from $q_0$ at $\bar q = 0$ to
\begin{align}
 \frac{\sum_{k=0}^K (1-{\rm e}^{-1})^k}{\sum_{k=0}^K \frac{(1-{\rm e}^{-1})^k}{q_k}} \cdot  {\rm e}^{-1} \nonumber
\end{align}
at $\bar q = 1$, it suffices to show that the above is less than or equal to ${\rm e}^{-1} $. In the meantime, \eqref{eq:mint} implies that the above is less than or equal to ${\rm e}^{-1} $. Therefore, \eqref{eq:FPE1} and \eqref{eq:FPE2} have a unique solution.
\end{proof}

Unlike Lemma \ref{lem:fixedpoint}, the forms of both \eqref{eq:FPE1} and \eqref{eq:FPE2} are fully exploited for the proof of Lemma \ref{lem:fixedpoint2}. Specifically, the fact that $\bar q ( 1 - \gamma)$ is an increasing function in $\bar q$ over the interval $[0,1]$ is used for the proof of Lemma \ref{lem:fixedpoint2}.

\begin{figure}[t!]
\centering
\centerline{\includegraphics[width=6cm]{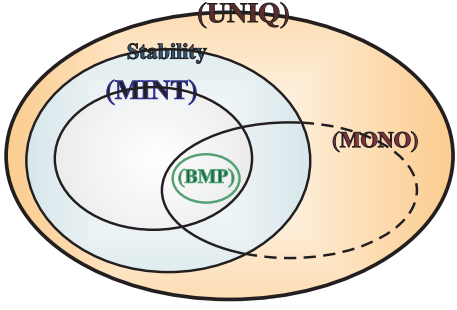}}
\caption{Logical relations between conditions.} \label{fig:venn}
\end{figure}

So far we have shown that there are two sufficient conditions, \eqref{eq:mono} and \eqref{eq:mint}, for the uniqueness of the equilibrium, \eqref{eq:uniq}, which is naturally a necessary condition for the global stability. We now show that one of them implies the global stability in the following theorem, which also {\it completes} the logical relations between \eqref{eq:mono}, \eqref{eq:uniq}, \eqref{eq:mint}, and the global stability, as shown in the Venn diagram in Fig. \ref{fig:venn}. Since it is not yet clear if there exists any case when \eqref{eq:mono} holds at the same time as the associated ODE is unstable, a part of the set \eqref{eq:mono} is depicted by the dashed line in Fig. \ref{fig:venn}.

It is remarkable that Lemma \ref{lem:fixedpoint2} is now rendered {\it obsolete} by the following theorem because the global stability of \eqref{eq:ODE} automatically implies \eqref{eq:uniq}, as clearly depicted in Fig. \ref{fig:venn}.

\begin{theorem}[Stability Condition]\label{th:meanfield}  \nextline
\eqref{eq:mint} implies the global stability of \eqref{eq:ODE}.
\end{theorem}
\begin{proof}
Because $\bar q(t)$ is bounded, there exist $\bar q^l$ and a sequence $\{ \tau_i \}$ such that $$\liminf_{t\to\infty} \bar q(t) = \bar q ^l, \quad \lim_{\tau_i \to\infty} \bar q(\tau_i) = \bar q ^l.$$  Since $\boldsymbol\phi(t)$ is a probability measure on a finite sample space $ \{ 0, \cdots,K \}$, $\boldsymbol\phi (t)$ is tight \cite{refBillConvergence}. 
Appealing to this, we can pick a convergent {\it sub}sequence $\{ t_i \}$ such that $\lim_{t_i \to \infty} \phi_k (t_i) = \phi_k (\infty) $ exists.

Defining $\nu(t) = \inf_{s\geq t} \bar q (s)$, we necessarily have $\nu(t) \leq \bar q(t)$, $\forall t\geq0$ and $\lim_{t\to\infty} \nu(t) = \bar q^l$. Consider the {\it degenerate} version of \eqref{eq:ODE} which has one additional equation with respect to $\frac{\ud \phi_0}{\ud t} ( t)$. By replacing $\bar q (t) $ with $ \nu(t)$, we get the following modified ODE:
\begin{align}
\frac{\ud \varphi_0}{\ud t} (t) & =
\nu(t)  {\rm e}^{- \nu(t) }  - {q_0} \varphi_0(t) + q_K \varphi_{K} (t) (1 - {\rm e}^{- \nu(t) }) , \nonumber\\
\frac{\ud \varphi_k}{\ud t} (t) & = q_{k-1} \varphi_{k-1} (t) (1 - {\rm e}^{- \nu(t) }) - q_k \varphi_k(t).\nonumber
\end{align}
Since $ \nu(t) $ becomes a constant for $t=\infty$, this ODE reduces to a linear ODE as $t\to \infty$ whose coefficient matrix takes the following form:
\begin{displaymath}
\left( \begin{array}{cccccc}
-q_0 & 0 & 0 &  \ldots & 0 & q_K \gamma^l  \\
q_0 \gamma^l  & -q_1 & 0 &  \ldots & 0 & 0 \\
0 & q_1 \gamma^l  & -q_2 &  \ldots & 0 & 0 \\
\vdots & \vdots & \vdots & \ddots & \vdots & \vdots \\
0 & 0 & 0 & \ldots & -q_{K-1} & 0 \\
0 & 0 & 0 & \ldots & q_{K-1} \gamma^l  & -q_K
\end{array} \right)
\end{displaymath}
where we used $ \gamma^l  \defeq \lim_{t \to\infty} (1 - {\rm e}^{- \nu(t) })$ for notational simplicity. Applying Gershgorin's circle theorem to the transpose of this coefficient matrix shows that all the eigenvalues are negative hence that $\varphi(t)$ converges as $t \to \infty$. Thus $\varphi(\infty)$ should satisfy
\begin{align}
&\textstyle {q_0}\varphi_0 (\infty)  = \bar q^l {\rm e}^{-\bar q^l} + q_K \varphi_K(\infty) \left( 1 - {\rm e}^{-\bar q^l} \right) ,\label{eq:varconv0}\\
&\textstyle q_k \varphi_k (\infty)  = q_{k-1} \varphi_{k-1}(\infty) \left( 1 - {\rm e}^{-\bar q^l} \right) ,\label{eq:varconvk}
\end{align}
for $k \in \{1,\cdots,K\}$ because $\lim_{t\to\infty} \nu(t) = \bar q^l$. Plugging \eqref{eq:varconvk} into \eqref{eq:varconv0} yields
\begin{equation}
q_k \varphi_k(\infty)  = \bar q^l ( 1 - {\rm e}^{-\bar q^l} )^k \left/ \sum_{j=0 }^K ( 1 - {\rm e}^{-\bar q^l} )^j \right. . \label{eq:varphisum}
\end{equation}

Suppose the initial condition $\varphi_k(0)=\phi_k(0)$, $\forall k\in \{0,\cdots,K\}$. We have the following equations from the modified ODE:
\begin{align}
& \varphi_0 (t)  = {\rm e}^{-q_0 t} \phi_0 (0) \nonumber \\&\textstyle + \int_{0}^{t} {\rm e}^{q_0 (s-t)} \left\{ \nu(s) {\rm e}^{-\nu(s)} + q_K \varphi_K (s) \left( 1 - {\rm e}^{-\nu(s)} \right) \right\} \ud s , \label{eq:varphi0}\\
& \varphi_k (t)  = {\rm e}^{-q_k t} \phi_k (0) \nonumber \\&\textstyle+  \int_{0}^{t} {\rm e}^{q_k (s-t)}   q_{k-1} \varphi_{k-1} (s) \left( 1 - {\rm e}^{-\nu(s)} \right)  \ud s , \label{eq:varphik}
\end{align}
where $k \in  \{1,\cdots,K\}$. First we have $ \nu(t) \leq 1 $ from the assumption
\eqref{eq:mint}. Since $1-{\rm e}^{-x}$ and $ x {\rm e}^{-x}$ terms in the above equations are increasing functions when $x \in [0,1]$ and $\nu(t) \leq \bar q(t)$, it can be checked by plugging \eqref{eq:varphik} into \eqref{eq:varphi0} $K$ times that $\varphi_0(t) \leq \phi_0(t)$ and hence $\varphi_k(t) \leq \phi_k(t)$, $\forall t\geq0$ and $\forall k\in \{0,\cdots,K\}$. That is, $\phi_k(t)$ is {\it lower-bounded} by $\varphi_k(t)$.

From \eqref{eq:varphisum} and the definition of the subsequence $\{ t_i \}$, we have the following relation:
$$\textstyle \sum_{k=0}^K q_k \varphi_k(\infty)  = \bar q^l = \sum_{k=0}^K q_k \phi_k(\infty). $$
where we recall $\phi_k(\infty)$ was defined as $\lim_{t_i \to \infty} \phi_k (t_i) = \phi_k (\infty) $. This result taken together with $\varphi_k(t) \leq \phi_k(t)$ proves $\varphi_k(\infty)=\phi_k(\infty)$, $\forall k \in \{0,\cdots,K\}$, and therefore, $\sum_{k=0}^K \varphi_k(\infty)=1$. Then it necessarily follows that $\bar q^l$ should satisfy \eqref{eq:FPE1} and \eqref{eq:FPE2} which have a unique solution by Lemma \ref{lem:fixedpoint2}. This implies $\bar q^l = \bar q$.

Note that we can also prove $\bar q^u = \bar q$ in a similar way by defining $\bar q^u$ and $\{ t_i \}$ such that $\limsup_{t\to\infty} \bar q(t) = \bar q ^u $, $\lim_{t_i \to\infty} \bar q(t_i) = \bar q ^u $ and $\lim_{t_i \to \infty} \phi_k (t_i) = \phi_k (\infty) $. This will show $\lim_{t\to \infty} \bar q (t) = \bar q$. That is, there is only one limit point for $\bar q(t)$.

Finally, we can pick a {\it new} sequence $\{ \tau_i \}$ such that $$\liminf_{t \to \infty} \phi_k(t) = \phi_k^l, \quad \lim_{\tau_i \to \infty} \phi_k(\tau_i) = \phi_k^l,$$ for all $k \in \{0,\cdots,K\}$. Using the fact $\lim_{t\to \infty} \bar q (t) = \bar q$, it can be easily proven that $\lim_{t\to \infty} \phi_k (t) = \phi_k$, $\forall k \in \{0,\cdots,K\}$, in a similar way. This establishes that $ ({\boldsymbol\phi}, \bar q, \gamma )$ is globally stable.
\end{proof}

\smallskip
\begin{remark}\label{rem:meanfield}
 This result gives an answer to the question raised in Section \ref{sec:intro} and {\bf justifies} the FPE approach used in \cite{refBianchi,refCarvalho,refGaretto,refKumar07,refKwak,refMedepalli,refRamaiyanMultistability,refSakuraiDelay} under a special scaling regime. That is, the decoupling assumption is validated in the asymptotic sense, as long as the scaled attempt rates are mild, \ie, \eqref{eq:mint}.

 The result of \cite[Theorem 5.4]{refBordenaveMF} implies that, for the case $K=\infty$, a set of strong conditions is required for the global stability of the ODE; a {\it monotonicity} condition along with a condition on attempt rate in backoff stage $k=0$, as shown in the condition \eqref{eq-bord} in Section \ref{sec:intro}. These strong conditions, designated also by \eqref{eq-bord} in Fig. \ref{fig:venn}, were proven to prevent wireless node to escape to infinite backoff stage. We can see that they correspond to a proper subset of the intersection of \eqref{eq:mint} and \eqref{eq:mono}. As compared with \cite[Theorem 5.4]{refBordenaveMF}, Theorem \ref{th:meanfield} is a stronger yet more practical argument due to finite $K$.

 As shown in Fig. \ref{fig:venn}, while the monotonicity \eqref{eq:mono} implies only the uniqueness \eqref{eq:uniq} which is not a decisive factor, \eqref{eq:mint} implies both \eqref{eq:uniq} and the global stability, assuring the asymptotic validity of the decoupling assumption. It is still {\it open} whether \eqref{eq:mono} implies the global stability or not.

 Informally, the proof of Theorem \ref{th:meanfield} follows from the fact that the solution $\boldsymbol\phi(t)$ cannot have more than one limit point. The {\it key observation} underlying its proof is that there exists a stable differential equation which becomes asymptotically linear as $t\to \infty$ at the same time as its solution $\varphi(t)$ lower-bounds $\boldsymbol\phi(t)$ such that $\boldsymbol\phi(t)$ is {\it squeezed} into $\boldsymbol\phi$ as $t \to \infty$.
\end{remark}
\smallskip

It is an intriguing fact that the above theorem may be restated in terms of $\gamma(t)$ rather than $q_k$, hence an alternative interpretation of the theorem: the ODE is globally stable if the collision probability $\gamma(t) \leq 1 -{\rm e}^{-1}=0.632$ for any initial condition ${\boldsymbol\phi}(0)= ( \phi_0 (0) ~ \cdots ~ \phi_K (0))^{\mathbf{T}}$. This interpretation means that if the collision probability is small enough, then the decoupling assumption is asymptotically valid, which appears to be in best agreement with our intuition.

As was mentioned at the beginning of this paper, there is only one intensity scaling regime $\epsilon(N) = 1/N$ which deserves to be analyzed because, under this regime, it is not clear whether the collision probability would converge to a unique equilibrium point and would stay around there forever. Though we have also shown that \eqref{eq:mint} is a sufficient condition for the asymptotic validity of the decoupling assumption, one may ask in return whether there exist {\it any} examples where \eqref{eq:mint} does not hold and the decoupling assumption is not asymptotically valid. Yes, there is. We will show in Section \ref{sec:oscillation} that the collision probability may oscillate between two values as time goes if \eqref{eq:mint} is violated.



\subsection{Achievable Throughput}\label{se:asympthroughput}

Recall that $L$ and $L_c$ denote the average duration of a successful packet transmission and the fixed duration of a collision, expressed in terms of backoff time-slot. The fixed overhead for each successful transmission is denoted by $L_o$. In what follows, we make a mild assumption that $ L_c \geq 1$ which means that the duration of a collision is no less than that of a single backoff time-slot. As we explained in Section \ref{sec:dcf}, $L_c$ is an RTS frame plus a DIFS, both of which is larger than a backoff time-slot in all versions of IEEE 802.11 MAC, regardless of the usage of the RTS/CTS mechanism.

Assuming that $\bar q(t) \to \bar q$ as $N$ tends to infinity, we can define the achievable throughput or alternatively the MFL of the aggregate throughput, as in \cite[Section 5]{refBordenaveSigmetrics08}:
\begin{align}\label{eq:asympth}
\Omega (\bar q) \defeq \frac{\probability_1 (\bar q) \cdot  L }{\probability_1 (\bar q) \cdot  (L +L_o) +  \probability_0 (\bar q) + \probability_c (\bar q) \cdot L_c }
\end{align}
where $\probability_1 (\bar q) \defeq \bar q {\rm e}^{-\bar q}$, $\probability_0 (\bar q) \defeq {\rm e}^{-\bar q}$, and $\probability_c (\bar q) \defeq 1 - \probability_1 (\bar q) - \probability_0 (\bar q)$ are the MFLs of the probabilities at each time-slot that only one node attempts transmission, none of the users attempts transmission, and at least two users attempt transmissions, respectively. Derivations of these MFLs are similar to that of \eqref{eq:gammaorig} and thus omitted.

Since \eqref{eq:asympth} holds on the condition that \eqref{eq:ODE} is {\it globally stable} such that $\lim_{t\to\infty} \bar q(t) = \bar q$, we can use \eqref{eq:asympth} so long as \eqref{eq:mint} holds. Then the result of Theorem \ref{th:meanfield} poses another question:

\vskip 3pt \noindent { \rmfamily\bfseries\mdseries\slshape ``Is there $q_k$ satisfying \eqref{eq:mint} and maximizing \eqref{eq:asympth} as well?''}
\vskip 3pt

Dividing the denominator of \eqref{eq:asympth} by its nominator, we can see that maximizing \eqref{eq:asympth} is equivalent to minimizing
$$ \frac{1}{\bar q} (1 - L_c) + \frac{{\rm e}^{\bar q}}{\bar q} L_c .$$
Differentiating this expression shows that the global maximum of \eqref{eq:asympth} is at the solution of the following equation:
\begin{align}\label{eq:qcondition}
\frac{1}{L_c} - 1 = (\bar q - 1) {\rm e}^{\bar q}
\end{align}
whose left-hand side is monotonically decreasing in $L_c$ over the domain $(0,\infty)$ and whose right-hand side is monotonically increasing in $\bar q $ over the same domain. Also both sides have the same range, \ie, $(-1,\infty)$. This implies, for each value of $L_c \in (0, \infty)$, there exists a unique solution to \eqref{eq:qcondition}, which is from now on denoted by $\bar q = \bar q^*$. If $L_c=1$, the solution is $\bar q^* = 1$. Putting these facts together, we can see that, if $L_c \geq 1$, there exists a solution $\bar q^* \leq 1$ to \eqref{eq:qcondition} which maximizes \eqref{eq:asympth}.

It is more important that $q_k$ satisfies \eqref{eq:mint} because \eqref{eq:mint} is a sufficient condition (and the only sufficient one we know) for the throughout equation \eqref{eq:asympth} to hold. To this aim, we show here that there are infinitely many constructions $q_k$ what satisfy \eqref{eq:mint} and \eqref{eq:mono} and maximize \eqref{eq:asympth} at the same time.
For $q_k = q_0 /m^k$ and $\bar q = \bar q^*$, plugging \eqref{eq:FPE2} into \eqref{eq:FPE1} yields:
\begin{align}\label{eq:maxfpe}
\frac{\bar q^*}{q_0} = \frac{\sum_{k=0}^K \left( 1 - {\rm e}^{-\bar q^*} \right)^k }{\sum_{k=0}^K \left(1 - {\rm e}^{-\bar q^*} \right)^k m^k }.
\end{align}
The right-hand side of \eqref{eq:maxfpe} is decreasing in $m \in (0,\infty)$. For given optimal solution $\bar q^*$, one can use \eqref{eq:maxfpe} to find $q_0$ and $m$ which satisfy \eqref{eq:mint} and maximize \eqref{eq:asympth} at the same time. For instance, in order to obtain nonincreasing $q_k$, one can simply set $q_0 = 1$ and compute $m$ from \eqref{eq:maxfpe} where $m \geq 1$ is warranted because the left-hand side of \eqref{eq:maxfpe} is no greater than $1$ and the right-hand side of \eqref{eq:maxfpe} decreases from $1$ at $m=1$ to $0$ at $m=\infty$.

To sum up, for every $q_0 \in [\bar q^*, 1]$ where $q^*$ is the solution to \eqref{eq:qcondition}, the construction $q_k = q_0/(m^*)^k$, where $m^*$ is the solution to \eqref{eq:maxfpe}, maximizes the aggregate throughput \eqref{eq:asympth} as well as guarantees the global stability of \eqref{eq:ODE}.

\section{Mean Field with Service Differentiation}\label{sec:mfdiff}

So far the discussion has centered on the homogeneous system where all nodes have the same parameter set. Now we turn to the heterogeneous case arising from the service differentiation mechanisms defined in 802.11e standard. In addition, a special kind of coupling caused by one of the mechanisms necessitates formulating a new ODE model.


\subsection{Prioritization Mechanisms}

Although three prioritization mechanisms are provided by enhanced distributed channel access (EDCA) functionality, one of which, called transmission opportunity (TXOP) \cite{refReduced}, exerts its influence only on time-slots when all nodes are freezed (See Section \ref{sec:dcf}), hence no need for making an analysis of it. The other two mechanisms are to differentiate per-class settings of
\begin{compactitem}
 \item contention window (CW),
  \item arbitration interframe space (AIFS).
\end{compactitem}

The first mechanism, CW differentiation, in the present context amounts to per-class setting of $q_0$ and $K$, on the assumption that  $q_k = q_0 /2^k$ for $k\in \{ 0, \cdots, K\}$. We extend this feature by allowing per-class setting of $K$ and $q_k$ for {\it any} $k\in \{ 0, \cdots, K\}$ for the sake of generality and notational aesthetics. Since CW differentiation implies that there are two or more classes, the corresponding system will be called {\bf heterogeneous}, whether the following differentiation is enabled or not.

The second, called AIFS differentiation, is to offer a {\it soft} non-preemptive prioritization to a certain class by holding back other classes from attempting transmissions for a few time-slots. This prioritization is effectuated by {\it idling} nodes for different durations, \ie, AIFS, after every transmission. In other words, AIFS differentiation {\it reserves} a few time-slots for high-priority classes.

The analysis here is presented for the case where there are two classes, \ie, Class $\rmH$ (high) and Class $\rmL$ (low), only to simplify the exposition, but can be extended to arbitrary number of classes. Let us call the time-slots reserved for Class $\rmH$ {\it reserved} slots, which will correspond to the superscript $\rmR$. We call the remaining slots following reserved slots {\it common} slots, corresponding to the superscript $\rmC$. Note that {\it both} Class $\rmH$ and Class $\rmL$ users can access the channel during common slots, whereas the backoff procedures of Class $\rmL$ users are suspended during reserved slots. The per-class parameters and occupancy measures are denoted by $q^\rmH_k$, $q^\rmL_k$, $K^\rmH$, $K^\rmL$, $\Phi^\rmH_k(t)$ and $\Phi^\rmL_k(t)$.

There are two kinds of couplings caused by the above-mentioned prioritization mechanisms.
\begin{compactitem}
 \item {\bf Inter-class coupling}: As compared with the analysis carried out in Section \ref{sec:revisited} where the stage evolution of nodes depends only on their own stage density, \ie, the occupancy measure $\boldsymbol\Phi(t) =  ( \Phi_0(t) ~ \cdots ~ \Phi_{K} (t))^{\mathbf{T}}$, the performance analysis of 802.11 in the presence of CW differentiation is complicated by the very fact that two-class users mutually interact with each other through $\boldsymbol\Phi^\rmH(t) \defeq ( \Phi^\rmH_0(t) ~ \cdots ~ \Phi^\rmH_{K^\rmH} (t))^{\mathbf{T}}$ and $\boldsymbol\Phi^\rmL(t) \defeq ( \Phi^\rmL_0(t) ~ \cdots ~ \Phi^\rmL_{K^\rmL} (t))^{\mathbf{T}}$. Fortunately, it turns out not very difficult to incorporate this complication into the ODE model in the previous section because we are simply dealing with two evolutions of the same kind.
  \item {\bf Coupling between two kinds of evolutions}: However, when it comes to AIFS differentiation, the issue is involved by the fact that the stage distribution of nodes in the previous time-slot affects the type of the current time-slot, and besides, the type of the current time-slot also affects the stage distribution of nodes in the next time-slot. That is, there are now two {\it different} kinds of evolutions, stage evolution of nodes and slot type evolution of time-slots, the latter of which adds a new type of state variable to the Markov chain model in \cite{refBianchi}. 
      An interesting point to note is that Sharma \etal \cite[Section IV]{refSharmaScaledMarkov} in a similar context also reckoned this difficulty though they have not solved it.
\end{compactitem}

Among many related works for modeling the AIFS differentiation, the works by Robinson and Randhawa \cite{refRobinson11e} and Ramaiyan \etal \cite{refRamaiyanMultistability} have taken the approach particularly relevant to our work. These works conducted their analyses under the assumption that per-class collision probabilities (or per-class average attempt rates) are constant over all time-slots. In what follows, we first analyze in Section \ref{sec:slottype} a Markov chain model of slot type evolution based on the intuition that slot type evolves {\it much faster} than per-class collision probabilities do and demonstrate in Section \ref{sec:extended_ode} that the result based on this analysis can be validated under the mean field regime by the result of \cite{refJYMF}, thereby a new variant of ODE model emerges.

\begin{figure*}[t!]
\centering
\centerline{\includegraphics[width=12cm]{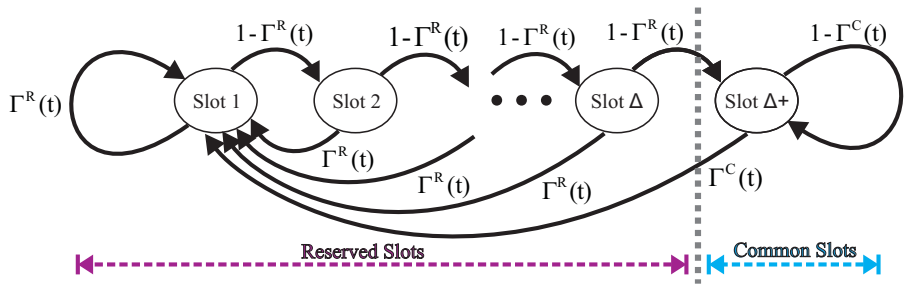}}
\caption{Evolution of slot type follows a nonhomogeneous Markov chain.} \label{fig:markov_aifs}
\end{figure*}

\subsection{Markov Model for the Evolution of Slot Type}\label{sec:slottype}

To avoid notational confusion, we use only the original occupancy measures in discrete time, \ie, $\Phi^\rmH_k(t)$ and $\Phi^\rmL_k(t)$ in this subsection. The MFLs of these variables will be defined in the next subsection. We first divide the population into two classes such that
$$N^\rmH + N^\rmL = N, ~\quad  \sigma^\rmH \defeq \frac{N^\rmH}{N}, \quad \sigma^\rmL \defeq \frac{N^\rmL}{N}.$$ Without loss of generality, the sets of nodes of Class $\rmH$ and Class $\rmL$ are denoted by $\N^\rmH \defeq \{ 1,\cdots, N^\rmH \} $ and $\N^\rmL \defeq \{ N^\rmH+1,\cdots, N \} $. Thus we define the occupancy measures as
\begin{align}\nonumber
\Phi^\rmH_k (t) \defeq \frac{1}{N} \sum_{n\in \N^\rmH} 1_{\{X_n (t) = k\} },~ \Phi^\rmL_k (t) \defeq \frac{1}{N} \sum_{n\in \N^\rmL} 1_{\{X_n (t) = k\} }
\end{align}
so that we have $\sigma^\rmH = \sum_{k=0}^{K^\rmH} \Phi^\rmH_k (t) $ and $\sigma^\rmL = \sum_{k=0}^{K^\rmL} \Phi^\rmL_k (t) $. Since there is no inter-class transition of users, $\sigma^{\rmH}$ and $\sigma^{\rmL}$ are constant and satisfy the relation $\sigma^{\rmH} + \sigma^{\rmL} = 1$.
In this setting, the probability that one or more nodes attempt transmission at time-slot $t$ of slot type $\rmR$ or $\rmC$ is given as follows:
\begin{align}
& \Gamma^\rmR(t)\! \defeq 1\! \! -\!  \prod_{k=0}^{K^\rmH} \left(1 - \epsilon(\!N\!) q^\rmH_k \right)^{N \cdot \Phi^\rmH_k( t)} , \label{eq:gammar_nominal}\\
& \Gamma^\rmC(t)\! \defeq 1\! \! -\!  \left(1-\Gamma^\rmR(t) \right) \prod_{k=0}^{K^\rmL} \left(1 - \epsilon(\!N\!) q^\rmL_k\right)^{N \cdot \Phi^\rmL_k( t)}. \label{eq:gammac_nominal}
\end{align}
Here we intentionally abuse the notation $\Gamma$ which is the same as the collision probability in \eqref{eq:gammaorig} because in the mean field limit the additional term $q_{k'}/N$ in \eqref{eq:gammaorig} vanishes, and thus the MFLs of the above equations and \eqref{eq:gammaorig} are much alike.

Now recall our assumption, $\epsilon(N)=1/N$, which was made in Section \ref{sec:valodemodel}. From the viewpoint of each individual node, we can describe AIFS differentiation by only three rules: (i) after any transmission attempt which is either successful or a failure, AIFS procedure is initialized, \ie, a counter value is set to zero; (ii) if the current time-slot is idle, the counter value is incremented by one; (iii) if the counter value reaches its designated per-class AIFS value, the node may attempt transmission with its per-stage probabilities, \ie, $q_{X_n(t)}^{\rmH} / N$ and $q_{X_n(t)}^{\rmL} / N$.

Denoting the difference of the two per-class AIFS values by $\Delta \geq 0$, we can see that the transition structure based on the aforementioned rules are illustrated by the {\it nonhomogeneous} Markov chain in Fig. \ref{fig:markov_aifs}, where we used the non-idle probabilities, \ie, $\Gamma^{\rmR}(t)$ and $\Gamma^{\rmC}(t)$, and the idle probabilities, \ie,  $1-\Gamma^{\rmR}(t)$ and $1-\Gamma^{\rmC}(t)$, as well. Here in Fig. \ref{fig:markov_aifs} reserved time-slots and common time-slots are respectively denoted by the notations `Slot 1'--`Slot $\Delta$' and `Slot $\Delta+$'. Note that $\Delta+$ means that, after any $\Delta$ or {\it more} consecutive {\it idle} backoff time-slots, the corresponding slot-type must be $\rmC$. It should be clear in Fig. \ref{fig:markov_aifs}  that not only slot-type but also the backoff stages of nodes, \ie, $\boldsymbol \Phi^{\rmH}(t)$ and $\boldsymbol \Phi^{\rmL}(t)$, are also changing over time-slots, hence $\Gamma^{\rmR}(t)$ and $\Gamma^{\rmC}(t)$ are.

The simplification in the analysis is made based on the following intuition that will be proven correct in Section \ref{sec:extended_ode}:

\smallskip
\noindent { \mdseries\slshape ``As population grows, the stage distribution (density) varies much slower than the type of time-slots.''}
\smallskip

 This observation follows essentially from the intensity scaling, $\epsilon(N)=1/N$, which leads to separation of time scales. This {\bf timescale decomposition} implies that slot type evolves on a relatively fast time scale, as compared with that of the evolution of occupancy measures. Formally speaking, the occupancy measures, $\Phi^\rmH_k(t)$ and $\Phi^\rmL_k(t)$, evolve at a rate of $\Theta(1/N)$ which ultimately vanishes as $N\to\infty$, whereas the probability that the slot-type changes for each time-slot does not vanish and remains strictly positive. Therefore, we can analyze the evolution of slot type {\bf as if} the occupancy measures were constant.
Working out the balance equations as if the nonhomogeneous Markov chain were homogeneous yields the following stationary distributions for each slot type:
\begin{align}\nonumber
\Pi^{\rmR} (t) & = \frac{ \sum_{i=0}^{\Delta-1} \left( 1- \Gamma^{\rmR}(t) \right)^i }{ \left\{ \sum_{i=0}^{\Delta-1} \left( 1- \Gamma^{\rmR}(t) \right)^i \right\} + \frac{\left( 1- \Gamma^{\rmR}(t) \right)^\Delta }{\Gamma^{\rmC}(t)} }, \\
\Pi^{\rmC} (t) & = \frac{ \frac{\left( 1- \Gamma^{\rmR}(t) \right)^\Delta }{\Gamma^{\rmC}(t)} }{ \left\{ \sum_{i=0}^{\Delta-1} \left( 1- \Gamma^{\rmR}(t) \right)^i \right\} + \frac{\left( 1- \Gamma^{\rmR}(t) \right)^\Delta }{\Gamma^{\rmC}(t)} } \nonumber
\end{align}
which satisfy $\Pi^{\rmR} (t) + \Pi^{\rmC} (t) \equiv 1$. Note however that in general it is impossible to derive the stationary distribution of  nonhomogeneous Markov chains where the transition probabilities are {\it time-varying}.

\subsection{Extended ODE Model with Prioritization Mechanisms}\label{sec:extended_ode}

The MFLs of $\Phi^\rmH_k(t)$ and $\Phi^\rmL_k(t)$ are denoted by $\phi^\rmH_k(t)$ and $\phi^\rmL_k(t)$ as in Section \ref{sec:valodemodel}.
As $N$ tends to infinity, by manipulation akin to \eqref{eq:collmf}, we can show that the collision probabilities for the different types of time-slots, \ie, \eqref{eq:gammar_nominal} and \eqref{eq:gammac_nominal}, become in the mean field regime:
\begin{align}\nonumber
& \gamma^{\rmR}(t) \defeq \lim_{N\to \infty} \Gamma^{\rmR}\left(t/\epsilon(N) \right) = 1 - {\rm e}^{- {\bar q^{\rmH}(t)}}, \\
& \gamma^{\rmC}(t) \defeq \lim_{N\to \infty} \Gamma^{\rmC}\left(t/\epsilon(N) \right) = 1 - {\rm e}^{- {\bar q^{\rmH}(t)} - {\bar q^{\rmL}(t)}} , \nonumber
 \end{align}
where the MFLs of the per-class average attempt rates are defined as
\begin{align}\nonumber
\bar q^{\rmH} (t) \defeq \sum_{k=0}^{K^{\rmH}} q^{\rmH}_k \phi^{\rmH}_k (t) \quad \textrm{and} \quad \bar q^{\rmL} (t) \defeq \sum_{k=0}^{K^{\rmL}} q^{\rmL}_k \phi^{\rmL}_k (t).
\end{align}

Now we derive the final extended ODE by using the $\gamma^\rmR(t)$ and $\gamma^\rmC(t)$. Let $\boldsymbol g^\rmR(t)$ and $\boldsymbol g^\rmC(t)$ denote the rates of change of $(\boldsymbol\phi^{\rmH}(t),\boldsymbol\phi^{\rmL}(t))$ when all time-slots are of slot type $\rmR$ or $\rmC$, respectively. As we did in Section \ref{sec:valodemodel}, we eliminate manifolds by using the equations $\phi^{\rmH}_0 (t) \equiv \sigma^{\rmH} - \sum_{k=1}^{K^{\rmH}} \phi^{\rmH}_k(t)$ and $\phi^{\rmL}_0 (t) \equiv \sigma^{\rmL} - \sum_{k=1}^{K^{\rmL}} \phi^{\rmL}_k(t)$ and hence we consider $(K^\rmH+K^\rmL)$-dimensional ODE. Since Class $\rmH$ users are allowed to attempt transmission only at time-slots of slot-type $\rmR$, whereas Class $\rmH$ users are allowed to do so at time-slots of any slot-type, the rates of change can be expressed as follows:
$$\boldsymbol g^\rmR(t) = \left( \begin{array}{c} q^{\rmH}_{K^{\rmH}-1} \phi^{\rmH}_{K^{\rmH}-1} (t) \gamma^{\rmR}(t) - q^{\rmH}_{K^{\rmH}} \phi^{\rmH}_{K^{\rmH}}(t) \\ \vdots \\ q^{\rmH}_{0} \phi^{\rmH}_{0} (t) \gamma^{\rmR}(t) - q^{\rmH}_1 \phi^{\rmH}_1(t) \\ \hdashline 0 \\ \vdots \\ 0 \end{array} \right), $$
$$\boldsymbol g^\rmC(t) = \left( \begin{array}{c} q^{\rmH}_{K^{\rmH}-1} \phi^{\rmH}_{K^{\rmH}-1} (t) \gamma^{\rmC}(t) - q^{\rmH}_{K^{\rmH}} \phi^{\rmH}_{K^{\rmH}}(t) \\ \vdots \\ q^{\rmH}_{0} \phi^{\rmH}_{0} (t) \gamma^{\rmC}(t) - q^{\rmH}_1 \phi^{\rmH}_1(t) \\ \hdashline q^{\rmL}_{K^{\rmL}-1} \phi^{\rmL}_{K^{\rmL}-1} (t) \gamma^{\rmC}(t) - q^{\rmL}_{K^{\rmL}} \phi^{\rmL}_{K^{\rmL}}(t) \\ \vdots \\ q^{\rmL}_{0} \phi^{\rmL}_{0} (t) \gamma^{\rmC}(t) - q^{\rmL}_1 \phi^{\rmL}_1(t) \end{array} \right). $$

It is remarkable that all the expressions intuitively derived based on the timescale decomposition in Section \ref{sec:slottype} can be formally {\bf justified} by the result of Bena\"im and Le Boudec \cite{refJYMF}. That is, the adopted simplification can be regarded as a natural consequence of the limiting regime. Formally speaking, we can apply \cite[Theorems 1 \& 2]{refJYMF}\footnotemark~to show that the resultant derivatives of $(\boldsymbol\phi^{\rmH}(t),\boldsymbol\phi^{\rmL}(t))$ become:
\begin{align}\label{eq:lcomprel}
\frac{\ud}{\ud t}\left( \begin{array}{c} \boldsymbol\phi^{\rmH}(t) \\ \boldsymbol\phi^{\rmL} (t) \end{array} \right) = \boldsymbol g^\rmR(t) \cdot \pi^{\rmR} (t) + \boldsymbol g^\rmC(t) \cdot \pi^{\rmC} (t)
\end{align}
\footnotetext{The corresponding assumptions can be easily checked.}
where $\pi^{\rmR} (t)$ and $\pi^{\rmC} (t)$ take the following forms
\begin{align}\nonumber
\pi^{\rmR} (t) & = \frac{ \sum_{i=0}^{\Delta-1} \left( 1- \gamma^{\rmR}(t) \right)^i }{ \left\{ \sum_{i=0}^{\Delta-1} \left( 1- \gamma^{\rmR}(t) \right)^i \right\} + \frac{\left( 1- \gamma^{\rmR}(t) \right)^\Delta }{\gamma^{\rmC}(t)} }, \\
\pi^{\rmC} (t) & = \frac{ \frac{\left( 1- \gamma^{\rmR}(t) \right)^\Delta }{\gamma^{\rmC}(t)} }{ \left\{ \sum_{i=0}^{\Delta-1} \left( 1- \gamma^{\rmR}(t) \right)^i \right\} + \frac{\left( 1- \gamma^{\rmR}(t) \right)^\Delta }{\gamma^{\rmC}(t)} } .\nonumber
\end{align}
This result has the implication that we can derive \eqref{eq:lcomprel} as a {\bf linear combination} of the MFL vectors, \ie, $\boldsymbol g^{\rmR}(t)$ and $\boldsymbol g^{\rmC}(t)$, with the coefficients, \ie, $\pi^{\rmR} (t)$ and $\pi^{\rmC} (t)$, which can be defined as $$\pi^{\rmR} (t) \defeq \lim_{N\to\infty} \Pi^{\rmR} (N t),\quad \pi^{\rmC} (t) \defeq \lim_{N\to\infty} \Pi^{\rmC} (Nt)  $$ where $\Pi^\rmR (\cdot)$ and $\Pi^\rmC (\cdot)$ are the stationary distributions we computed from Fig. \ref{fig:markov_aifs} in Section \ref{sec:slottype} as if the nonhomogeneous Markov chain were homogeneous.

Finally, after some manipulation of \eqref{eq:lcomprel}, we have the following enhanced ordinary differential equation:
\begin{align}
\frac{\ud \phi^{\rmH}_k}{\ud t} (t) & = q^{\rmH}_{k-1} \phi^{\rmH}_{k-1} (t) \gamma^{\rmH}(t) - q^{\rmH}_k \phi^{\rmH}_k(t),\label{eq:EODEH} \\
\frac{\ud \phi^{\rmL}_k}{\ud t} (t) & = \pi^{\rmC} (t) \left\{ q^{\rmL}_{k-1} \phi^{\rmL}_{k-1} (t) \gamma^{\rmC}(t) - q^{\rmL}_k \phi^{\rmL}_k(t) \right\} ,\label{eq:EODEL}
\end{align}
where \eqref{eq:EODEH} and \eqref{eq:EODEL} respectively hold for $k\in \{1,\cdots,K^{\rmH} \}$ and $k\in \{1,\cdots,K^{\rmL} \}$. Here we use the following shorthand notation:
 \begin{align}\nonumber
 \gamma^{\rmH}(t) = \pi^{\rmR} (t)\gamma^{\rmR}(t) + \pi^{\rmC} (t)\gamma^{\rmC}(t)  \nonumber
 \end{align}
whose form is obvious from \eqref{eq:lcomprel}.

In the stationary regime, we can get the following fixed point equation:
\begin{align}\label{eq:EFPE1}
\bar q^{\rmH}  & =  \sigma^{\rmH} \frac{\sum_{k=0}^{K^{\rmH}} \left( \gamma^{\rmH} \right)^k}{\sum_{k=0}^{K^{\rmH}} \frac{ \left( \gamma^{\rmH} \right)^k }{q_k^{\rmH}}}, \\
 \bar q^{\rmL} & =  \sigma^{\rmL} \frac{\sum_{k=0}^{K^{\rmL}} \left( \gamma^{\rmC} \right)^k}{\sum_{k=0}^{K^{\rmL}} \frac{ \left( \gamma^{\rmC} \right)^k }{q_k^{\rmL}}}, \label{eq:EFPE2}\\
\gamma^{\rmH}  & = \pi^{\rmR} \left( 1 -  {\rm e}^{- {\bar q^{\rmH}} } \right) + \pi^{\rmC} \left( 1 - {\rm e}^{- {\bar q^{\rmH}} - {\bar q^{\rmL}}} \right) , \label{eq:EFPE3} \\
\gamma^{\rmC}  & = 1 - {\rm e}^{- {\bar q^{\rmH}} - {\bar q^{\rmL}}}. \label{eq:EFPE4}
\end{align}

\smallskip
\begin{remark}\label{rem:eode}
 It is remarkable that the extended ODE model laid out in \eqref{eq:EODEH} and \eqref{eq:EODEL} {\it encompasses} the homogeneous system in Section \ref{sec:justification}, and the heterogeneous system in Section \ref{sec:mfdiff} as well, which has the two prioritization functionalities.

 For instance, if $\Delta=\infty$, we have $\pi^\rmR (t) = 1$ and the ODE model reduces to the homogeneous system \eqref{eq:ODE}. On the other hand, if $\Delta = 0$, we have $\pi^\rmC (t) = 1$ and the ODE model reduces to a purely heterogeneous system, implying that the AIFS differentiation is disabled.

 What is the most surprising is that the FPE \eqref{eq:EFPE1}-\eqref{eq:EFPE4} coincides with that proposed in \cite[Section VI]{refRamaiyanMultistability}, which was derived rather intuitively.
\end{remark}
\smallskip

In the following, we introduce three new conditions akin to those in Section \ref{sec:valodemodel}.
\begin{align}
q^\rmH_k \mbox{ and } q^\rmL_k  \mbox{ are nonincreasing in }k \mbox{,}
 \label{eq:emono}
\end{align}
\begin{align}
\mbox{\eqref{eq:EFPE1}-\eqref{eq:EFPE4} has a unique solution,} \label{eq:euniq}
\end{align}
\begin{align}
 q^\rmH_k \leq 1 \mbox{ and }  q^\rmL_k \leq 1, ~\forall k.
 \label{eq:emint}
\end{align}
By adopting the above conditions we present two lemmas.
\begin{lem}[Monotonicity Implies Uniqueness]\label{lem:efixedpoint} \nextline
\eqref{eq:emono} implies \eqref{eq:euniq}.
\end{lem}
\begin{lem}[Mild Intensity Implies Uniqueness]\label{lem:efixedpoint2} \nextline
\eqref{eq:emint} implies \eqref{eq:euniq}.
\end{lem}

Proofs of the above two lemmas are in Appendix. It is of importance to note that these lemmas are of even {\it greater} generality because they hold for all $\Delta \geq 0$ and $\Delta = \infty$ as well, implying that Lemmas \ref{lem:fixedpoint} and \ref{lem:fixedpoint2} respectively correspond to the special cases of Lemmas \ref{lem:efixedpoint} and \ref{lem:efixedpoint2}, \ie, the case $\Delta = \infty$.

We are not able to prove the equivalent of Theorem \ref{th:meanfield} for the case where there are more than one class due to the inter-class coupling arising from CW differentiation. This coupling makes it technically challenging to find a stable ODE, which would bound the solution of the ODE as in the proof of Theorem \ref{th:meanfield}. In the meantime, the other coupling induced by AIFS differentiation does not seem to cause a major technical difficulty. As of now, we have to be content with having stated the problem precisely with its inherent technical difficulty.

\section{Selected Counterexamples}\label{sec:examples}

Before proceeding to selected examples, the gap between the ODE model and the backoff processes in 802.11 must be bridged. This gap emerged right on applying the intensity scaling in Section \ref{sec:valodemodel}. The scaling relation $p_k = q_k/N$ suggests to us that  replacing $q_k$ by $ N p_k$ should yield a reasonable approximation if $p_k$ is small. Plugging this approximation and removing the time acceleration from \eqref{eq:ODE}, we have
\begin{align}
\frac{\ud \phi_k}{\ud t} (t) & = p_{k-1} \phi_{k-1} (t) \gamma(t) - p_k \phi_k(t) \label{eq:ODE'}
\end{align}
where $\gamma(t) \defeq 1 - {\rm e}^{-N \bar p(t)}$ and $\bar p (t) \defeq \sum_{k=0}^{K} p_k \phi_k (t)$.

In this section, we provide two counterexmaples which will demonstrate versatility of the ODE model. In particular, two goals of the simulation are as follows:
\begin{compactitem}
 \item  By comparing the trajectories of the ODE model \eqref{eq:ODE'} with the simulation result of the corresponding Discrete Time Markov Chain (DTMC), we show that the simplistic ODE model is accurate enough to provide us insights into the formidably complex DTMC.
 \item For the case the stability condition \eqref{eq:mono} is violated, the examples illustrate two major unstable behavior patterns of the system.
\end{compactitem}

\begin{figure}[t!]
\begin{center}
  \subfigure[$ f(\gamma) $ versus $\gamma $]{\label{fig:ex1root}
        \includegraphics[width=7.9cm]{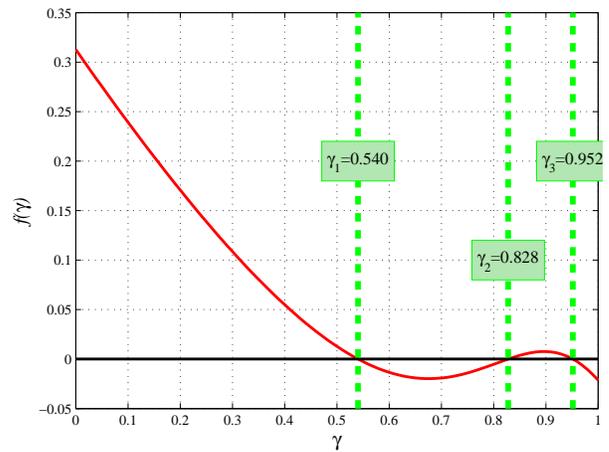}}
 \subfigure[Short-term average collision probability vs. backoff time-slots]{\label{fig:ex1instcol}
        \includegraphics[width=7.9cm]{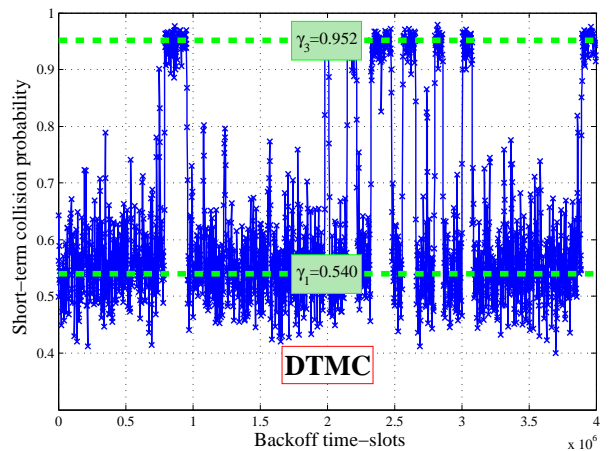}}
  \subfigure[Short-term average occupancy measure in backoff stages, $(\phi_0(t), \phi_{1}(t))$, $(\phi_0(t), \phi_{2}(t))$ and $(\phi_0(t), \phi_{3}(t))$. {\itshape Stars ($\star$) and circles ($\circ$)}: mean field limits; {\itshape dots}: DTMC simulation.]{\label{fig:ex1instocc}
        \includegraphics[width=7.9cm]{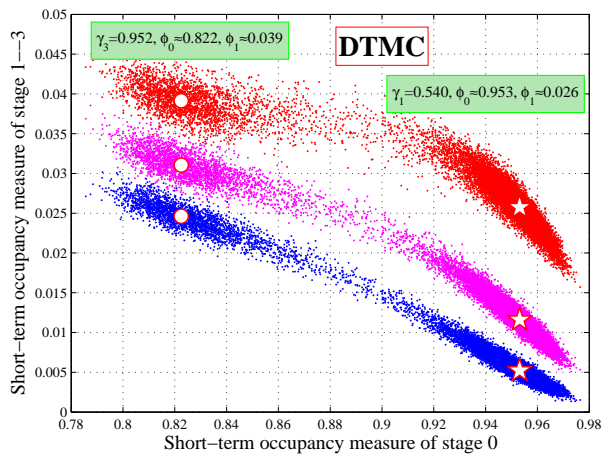}}
\caption{Bistabilty Example: There are three solutions to the fixed point equation, two of which ($\gamma_1$ and $\gamma_3$) are stable and the other one ($\gamma_2$) is unstable. Short-term average statistics measured for each 2000 backoff time-slots suggest bistability.} \label{fig:ex1}
\end{center}
\end{figure}

 Note that all Markov chains used for the simulation are ergodic: statistically speaking, the two systems will forget their initial states after evolving for a long enough time. To this aim, we have run each of the DTMC simulations for $120,000,000$ backoff time-slots. It is remarkable that we must run the DTMC simulations for a very long duration because the DTMC in Section \ref{sec:multistability} exhibits a phenomenon called bistability, which can be observed by running simulations for a relatively long time (\eg, see Fig. \ref{fig:ex1instcol}).

 To obtain the short-term average statistics, the entire duration of each simulation is divided into disjoint intervals of $2,000$ time-slots and each short-term average data point was calculated over one of the disjoint intervals. We assume that all wireless nodes are in backoff stage $0$ at the initial time-slot.

\subsection{Example 1: Multistability}\label{sec:multistability}

Consider the {\bf homogeneous} system \eqref{eq:ODE'}. Plugging \eqref{eq:NFPE1} into \eqref{eq:NFPE2} yields:
\begin{align}
f(\gamma) \defeq &  1 - {\rm exp}\left({- N \frac{\sum_{k=0}^{K} \gamma^k}{\sum_{k=0}^{K} \frac{\gamma^k}{p_k}}} \right) - \gamma = 0 \label{eq:simplefpe}
\end{align}
which is a function of only $\gamma$. Consider the following multistability example where there are $N=1200$ nodes and $K+1=13$ backoff stages. The attempt probability at each backoff stage $p_k$ is $$ (p_0, p_1, \cdots, p_{12}) = \left( \frac{1}{3200}, \frac{1}{160}, \frac{m}{160}, \cdots, \frac{m^{11}}{160}  \right) $$ where $m=6/5=1.2$. The roots of \eqref{eq:simplefpe} can be computed from Fig. \ref{fig:ex1root} as
$$ (\gamma_1 , \gamma_2, \gamma_3 ) = (0.540, 0.828, 0.952). $$

The instantaneous collision probability for each 2000 backoff time-slots is shown in Fig. \ref{fig:ex1instcol}, which tends to concentrate around $\gamma_1 = 0.540$ and $\gamma_3 = 0.952$. Note that the average collision probability for the entire duration of the simulation is $0.832$ that is neither $\gamma_1$ nor $\gamma_3$.
  Recall that $\phi_k(t)$ denotes the fraction of nodes in backoff stage $k$. Fig. \ref{fig:ex1instocc} shows the short-term average of the fraction of nodes in backoff stage $k \in \{ 1, 2,3 \}$ versus that in backoff stage $0$. From the top to the bottom, the short-term occupancy measures of stages $1-3$ are shown in order, where the two kinds of markers, \ie, circle ($\circ$) and star ($\star$), stand for the occupancy measures at two equilibriums, $\gamma_3$ and $\gamma_1$, which are computed from \eqref{eq:ODE'}. The {\it bistability} of this system is precisely predicted from either two modes of behavior of \eqref{eq:ODE'} or the eigenvalues of Jacobian matrices at the three equilibrium points.

\subsection{Example 2: Stable Oscillation}\label{sec:oscillation}

\begin{figure}[h!]
\begin{center}
  \subfigure[$ f(\gamma) $ versus $\gamma $]{\label{fig:ex2root}
        \includegraphics[width=7.9cm]{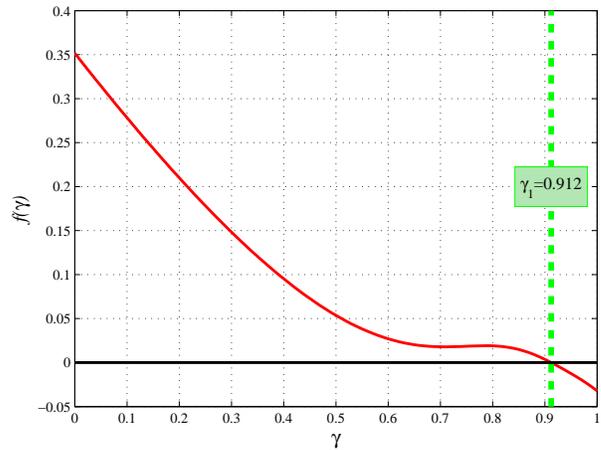}}
 \subfigure[Short-term average collision probability vs. backoff time-slots]{\label{fig:ex2instcol}
        \includegraphics[width=7.9cm]{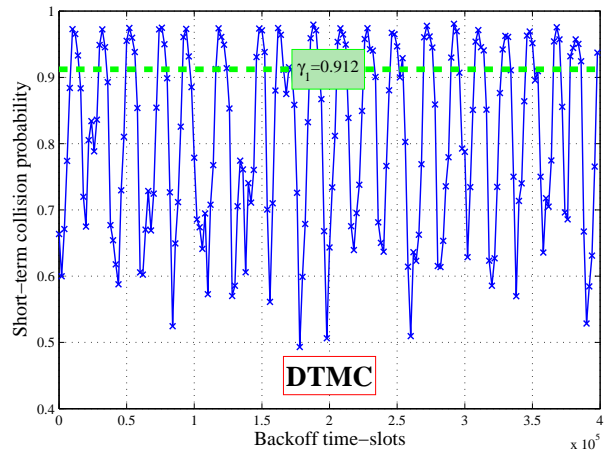}}
  \subfigure[Short-term average occupancy measure in backoff stages, $(\phi_0(t), \phi_{1}(t))$ and $(\phi_0(t), \phi_{17}(t))$. {\itshape Solid lines and stars ($\star$)}: mean field limits; {\itshape dots}: DTMC simulation.]{\label{fig:ex2instocc}
        \includegraphics[width=7.9cm]{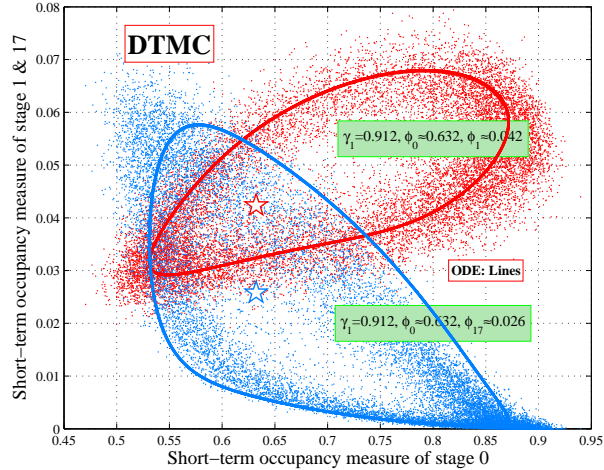}}
\caption{Oscillation Example: There is a unique solution ($\gamma_1$) to the fixed point equation but the decoupling assumption does {\bf\itshape not} hold in the asymptotic sense. Short-term average statistics measured for each 2000 backoff time-slots suggest stable oscillation around the unique equilibrium.} \label{fig:ex2}
\end{center}
\end{figure}

We have managed to discover a rare example by delving into the {\bf heterogeneous} system, without AIFS differentiation, \ie, $\Delta = 0$, which in turn leads to $\pi^\rmR = 0$ and $\pi^\rmC = 1$. Suppose there are two classes $\rmH$ and $\rmL$ such that population of each class is $N^\rmH=N^\rmL=640$. The numbers of backoff stages are assumed to be equal, \ie, $K^\rmH + 1 = K^\rmL + 1=  21$. The attempt probability at each backoff stage is:
$$ (p^\rmH_0, p^\rmH_1, \cdots, p^\rmH_{20}) = \left( \frac{1}{2400}, \frac{1}{480}, \frac{m}{40}, \cdots, \frac{m^{19}}{40}  \right) $$
$$ (p^\rmL_0, p^\rmL_1, \cdots, p^\rmL_{20}) = \left( \frac{1}{3840}, \frac{1}{64}, \frac{1}{64}, \cdots, \frac{1}{64}  \right) $$
where $m=4/5$. It is easy to verify that the corresponding fixed point equation takes the following form:
\begin{align}
f(\gamma) \defeq & = 1 - \prod_{\rmX \in \{\rmH,\rmL\}} {\rm exp}\left(- N^\rmX \frac{\sum_{k=0}^{K^\rmX} \gamma^k}{\sum_{k=0}^{K^\rmX} \frac{\gamma^k}{p^\rmX_k}}  \right) - \gamma = 0 \nonumber 
\end{align}
which has the following {\bf unique} solution as shown in Fig. \ref{fig:ex2root}:
$$ \gamma^\rmH = \gamma^\rmR = \gamma^\rmC = \gamma_1 = 0.912. $$

Since there is only one solution, one might be much inclined to hazard the conjecture by Bianchi \etal \cite{refBianchi,refKumar07} that the collision probability is approximately $\gamma_1$. However, there is a stable limit cycle around this equilibrium. In other words, the oscillation is stable, \ie, not transient but lasting forever. The event-average collision probability obtained through simulations is $0.869$ which is less than $\gamma^\rmH $ or $\gamma^\rmC$.

We can see from Fig. \ref{fig:ex2instcol} that, unlike the previous example, the trajectory of instantaneous collision probability forms almost periodic oscillation and does not tend to concentrate around the unique equilibrium $\gamma_1$.
Though the oscillation is not deterministic but stochastic, it clearly persists as time goes to infinity. The period of the oscillation empirically can be computed from Fig. \ref{fig:ex2instcol} as between $19000$ and $20000$ time-slots. The oscillation and its period are exactly predicted from the trajectories of the ODE model (sold lines) as shown in Fig. \ref{fig:ex2instocc}. The unstability of $\gamma_1$ can be decided by the eigenvalues of the corresponding Jacobian matrix.

The decoupling assumption does not hold in the asymptotic sense; in contrast, nodes are coupled by the oscillations of the occupancy measure, an emerging property of the system dynamics.

\section{Concluding Remarks with a Conjecture}\label{sec:conclusion}

Since it is axiomatic that the fixed point equation (FPE), called Bianchi's formula, must have a unique solution in order to provide an approximation, there has been a speculation that the uniqueness of the solution might assure the validity of the FPE, which has been the main subject of previous approaches by Kumar \etal \cite{refKumar07}, and Ramaiyan \etal \cite{refRamaiyanMultistability}. One counterexample in our paper has shown that this speculation is not always true, putting another emphasis on asymptotic validation of the decoupling assumption which underlies the formula.

Thanks to recent advances in mean field theory \cite{refBordenaveMF,refJYMF} and also \cite{refSharmaScaledMarkov}, we have analyzed the validity of the FPE by determining the stability of an ordinary differential equation (ODE). In the course of establishing stability, we obtained an illuminating insight that not only monotonicity but also mildness of scaled attempt rate guarantees the uniqueness of the equilibrium, which made the logical relations between them clear. Paradoxically, the mathematical formalism of mean field theory presented us a {\it succinct} stability condition \eqref{eq:mint}, whose main implication is as follows: to achieve perfect decoupling between nodes as population $N$ grows, in addition to reducing the attempt probability at $k$th backoff to $q_k /N$, we need to {\it further} diminish the node activity such that the scaled attempt rate satisfies $q_k \leq 1$. The existence of such an upper bound appears to be in best agreement with our usual intuition. We also have shown that there are infinitely many constructions of $q_k$ which maximize the aggregate throughput as well as satisfy \eqref{eq:mint} and \eqref{eq:mono}, hence the condition \eqref{eq:mint} is practical as well.

Though an EDCA prioritization mechanism causes a new type of coupling between the evolutions of per-class remaining idle times and backoff stages, which has been an intricate complication \cite{refSharmaScaledMarkov}, another penetration, also formalized by mean field argument, has led us to an extended form of an ODE model spinning off a generalized FPE as well.

Lastly, we conjecture that \eqref{eq:mint} implies the global stability of \eqref{eq:EODEH} and \eqref{eq:EODEL} as well, as observed in our exhaustive simulations. We believe that it is provable with a Lyapunov function though the form of which is unknown yet. Although theoretical support to this conjecture is not available, we hope the discussion can introduce the challenging side of the open stability problem.

\appendix
\section{Appendix: Proofs}

\subsection{Alternative Proof of Lemma \ref{lem:fixedpoint}}\label{sec:proofpropfixed1}

 We show the existence and uniqueness of the equilibrium point. Differentiating the right-hand side of \eqref{eq:FPE1} with respect to $\bar q$, we can see that the following equation determines the sign of the derivative.
\begin{align*}
\textstyle\delta_K  & \defeq \sum_{k =0}^{K} k \gamma^{k-1} \left( \sum_{j =0}^K \frac{ \gamma^j}{q_j} \right) - \sum_{j = 0}^K \gamma^j \left( \sum_{k =0 }^K  \frac{k  \gamma^{k-1}}{q_k} \right)\\
 &= \textstyle\sum_{k =0}^K \sum_{j =0}^K \gamma^{k+j-1} \left( \frac{k}{q_j} - \frac{k}{q_k} \right).
\end{align*}
Consider a proper subsum, $\delta_\kappa$, which can obtained by replacing $K$ with $\kappa \in \{1,\cdots,K-1 \}$. Recall that $q_0 \geq q_1$ by the assumption; then it is easy to see that $\delta_1 \leq 0 $ is true. Now suppose $\delta_\kappa$ is zero or negative. We show $\delta_{\kappa+1} \leq 0$ if $\delta_\kappa \leq 0$. Rearranging terms of $\delta_{\kappa+1}$, it is not difficult to obtain:
\begin{align*}
\textstyle\delta_{\kappa+1} & = \textstyle\delta_\kappa \textstyle+\left[ \sum_{i=0}^{\kappa} \gamma^{\kappa+i} \left( \kappa+1-i \right) \left( \frac{1}{q_i} - \frac{1}{q_{\kappa+1}} \right) \right] 
\end{align*}
where the second term on the right-hand side is zero or negative as $ q_i $ is nonincreasing for $i \in \{ 0,\cdots,K \}$. As  $\delta(K)$ is zero or negative, we can conclude that the right-hand side of \eqref{eq:FPE1} is a nonincreasing function which is positive and converges to $(K+1)/\sum_{k=0}^K q_k^{-1} $ at $\bar p =\infty$. This conclusion taken together with the fact that the left-hand side of \eqref{eq:FPE1} is an identical function from $[0,\infty)$ to $[0,\infty)$ proves that there exists a unique equilibrium point $\bar q$.

\subsection{Proof of Lemma \ref{lem:efixedpoint}}\label{sec:proofepropfixed}

First, we note from \eqref{eq:emono} that the right-hand sides of \eqref{eq:EFPE1} and \eqref{eq:EFPE2} are nonincreasing in $\gamma^\rmH$ and $\gamma^\rmC$, respectively. The proof of this fact is almost identical to that of Lemma \ref{lem:fixedpoint}.

Assume that there are two solutions $( \bar q^{\rmH}, \bar q^{\rmL})$ and $(\acute{\bar q}^{\rmH}, \acute{\bar q}^{\rmL})$ of the fixed point equation \eqref{eq:EFPE1}-\eqref{eq:EFPE4} and $\acute{\bar q}^{\rmH} \geq \bar q^{\rmH}$, without loss of generality. If we assume that $\acute{\bar q}^{\rmL} \geq \bar q^{\rmL}$, it follows from \eqref{eq:EFPE4} that $\acute\gamma^{\rmC} \geq \gamma^{\rmC}$. Because the right-hand side of \eqref{eq:EFPE2} is nonincreasing in  $\gamma^\rmC$, we must have $\acute{\bar q}^{\rmL} \leq \bar q^{\rmL}$ and hence $\acute\gamma^{\rmC} \geq \gamma^{\rmC}$. Now we have shown by contradiction that $\acute{\bar q}^{\rmH} \geq \bar q^{\rmH}$ implies $\acute{\bar q}^{\rmL} \leq \bar q^{\rmL}$ and $\acute\gamma^{\rmC} \geq \gamma^{\rmC}$.

Moreover, we can rewrite \eqref{eq:EFPE3} in the following form:
\begin{align}\nonumber
\gamma^{\rmH} & = \frac{ \left( 1- \gamma^\rmR \right)^\Delta + \gamma^\rmR \sum_{i=0}^{\Delta-1} \left( 1- \gamma^{\rmR} \right)^i }{ \left\{ \sum_{i=0}^{\Delta-1} \left( 1- \gamma^{\rmR} \right)^i \right\} + \frac{\left( 1- \gamma^{\rmR} \right)^\Delta }{\gamma^{\rmC}} } \\ & = \frac{ 1 }{ \left\{ \sum_{i=0}^{\Delta-1} \left( 1- \gamma^{\rmR} \right)^i \right\} + \frac{\left( 1- \gamma^{\rmR} \right)^\Delta }{\gamma^{\rmC}} } \label{eq:temppir}
\end{align}
where the second equality can be easily verified. As $\gamma^\rmR = 1 - {\rm e}^{-\bar q^\rmH}  $ is increasing in $\bar q^\rmH$, $\acute{\bar q}^{\rmH} \geq \bar q^{\rmH}$ implies $\acute\gamma^{\rmC} \geq \gamma^{\rmC}$ and $\acute\gamma^{\rmR} \geq \gamma^{\rmR}$. Combining these with the fact that \eqref{eq:temppir} is increasing in $\gamma^\rmR$ and $\gamma^\rmC$, we can establish that $\acute{\bar q}^{\rmH} \geq \bar q^{\rmH}$ implies $\acute\gamma^{\rmH} \geq \gamma^{\rmH}$. On the other hand, since the right-hand side of \eqref{eq:EFPE1} is nonincreasing in $\gamma^\rmH$, the inequality $\acute\gamma^{\rmH} \geq \gamma^{\rmH}$ must imply $\acute{\bar q}^{\rmH} \leq \bar q^{\rmH}$.

In conclusion, if we assume $\acute{\bar q}^{\rmH} \geq \bar q^{\rmH}$, we have $\acute{\bar q}^{\rmH} \leq \bar q^{\rmH}$, which implies that $\acute{\bar q}^{\rmH} = \bar q^{\rmH}$. Then it automatically follows that $\acute{\bar q}^{\rmL} = \bar q^{\rmL}$, $\acute\gamma^{\rmH} = \gamma^{\rmH}$, and $\acute\gamma^{\rmL} = \gamma^{\rmL}$.

We have yet to establish the existence of the solution. We first note that the left-hand sides of \eqref{eq:EFPE1} and \eqref{eq:EFPE2} are identical functions of $\bar q^\rmH$ and $\bar q^\rmL$, respectively, from $[0,\infty)$ to $[0,\infty)$. Because \eqref{eq:EFPE4} is increasing in $\bar q^\rmL$, for each fixed $\bar q^\rmH$, the right-hand side of \eqref{eq:EFPE2} is a positive nonincreasing function of $\bar q^\rmL$ by the proof of Lemma \ref{lem:fixedpoint}. Likewise, as \eqref{eq:EFPE3} is increasing in $\bar q^\rmH$ for each fixed $\bar q^\rmL$, the right-hand side of \eqref{eq:EFPE1} is a positive nonincreasing function of $\bar q^\rmH$ by the proof of Lemma \ref{lem:fixedpoint}. This completes the proof.

\subsection{Proof of Lemma \ref{lem:efixedpoint2}}\label{sec:proofepropfixed2}

Multiplying both sides of \eqref{eq:EFPE1} and \eqref{eq:EFPE2} respectively by $(1- \gamma^\rmH)$ and $(1-\gamma^\rmC)$ yields the following equations:
\begin{align}\label{eq:EFPE1'}
\bar q^{\rmH} (1- \gamma^\rmH) & =  \sigma^{\rmH} \frac{\sum_{k=0}^{K^{\rmH}} \left( \gamma^{\rmH} \right)^k}{\sum_{k=0}^{K^{\rmH}} \frac{ \left( \gamma^{\rmH} \right)^k }{q_k^{\rmH}}} \cdot (1- \gamma^\rmH), \\
 \bar q^{\rmL} (1- \gamma^\rmC) & =  \sigma^{\rmL} \frac{\sum_{k=0}^{K^{\rmL}} \left( \gamma^{\rmC} \right)^k}{\sum_{k=0}^{K^{\rmL}} \frac{ \left( \gamma^{\rmC} \right)^k }{q_k^{\rmL}}} \cdot (1- \gamma^\rmC) \label{eq:EFPE2'}.
\end{align}
The proof is similar to that of Lemma \ref{lem:efixedpoint} except that:
\begin{asparaenum}[\rmfamily(i)]
\item We use the fixed point equation \eqref{eq:EFPE1'}, \eqref{eq:EFPE2'}, \eqref{eq:EFPE3} and \eqref{eq:EFPE4}.
 \item We note from Lemma \ref{lem:fixedpoint2} that the right-hand sides of \eqref{eq:EFPE1'} and \eqref{eq:EFPE2'} are decreasing respectively in $\bar q^{\rmH}$ and $\bar q^{\rmL}$, and less than or equal to the left-hand sides of \eqref{eq:EFPE1'} and \eqref{eq:EFPE2'} respectively at $\bar q^{\rmH}=1$ and $\bar q^{\rmL}=1$.
\end{asparaenum}

 To complete the proof, it is sufficient to show that the left-hand sides of \eqref{eq:EFPE1'} and \eqref{eq:EFPE2'} are increasing respectively in $\bar q^{\rmH}$ and $\bar q^{\rmL}$. It follows from the proof of Lemma \ref{lem:fixedpoint2} that \eqref{eq:emint} implies $\bar q^\rmH \leq 1$ and $\bar q^\rmL \leq 1$. It is also obvious from the form of $\bar q^{\rmL} (1- \gamma^\rmC) = \bar q^{\rmL} {\rm e}^{-\bar q^\rmH - \bar q^\rmL}$ that the left-hand side of \eqref{eq:EFPE2'} in increasing in $\bar q^\rmL \in [0,1]$.

To sum up again, it is now enough to show that the left-hand side of \eqref{eq:EFPE1'} is increasing in $\bar q^\rmH \in [0,1]$. To establish this, we rewrite \eqref{eq:EFPE3} in a compact form
\begin{align}
\gamma^{\rmH}  = 1 - \left\{ \frac{ {\rm e}^{-\bar q^{\rmH}} h(\bar q^{\rmH} , \bar q^{\rmL})    }{ h(\bar q^{\rmH} , \bar q^{\rmL}) + 1 } + \frac{ {\rm e}^{-\bar q^{\rmH} -\bar q^\rmL}  }{ h(\bar q^{\rmH} , \bar q^{\rmL}) + 1 } \right\} \nonumber
\end{align}
where $ h(\bar q^{\rmH} , \bar q^{\rmL}) \defeq \left( {{\rm e}^{\bar q^{\rmH} \Delta}-1} \right) \cdot \frac{1 - {\rm e}^{- \bar q^{\rmH} - {\bar q^{\rmL}}}}{ 1-{\rm e}^{-\bar q^{\rmH}} } $. Differentiating $\bar q^{\rmH}(1-\gamma^\rmH)$ with respect to $\bar q^{\rmH}$ yields
\begin{align}
(1 - \bar q^{\rmH} ) \frac{{\rm e}^{-\bar q^{\rmH}}h + {\rm e}^{-\bar q^{\rmH} - \bar q^\rmL }}{h+1}  + \bar q^{\rmH} \frac{{\rm e}^{-\bar q^{\rmH}} - {\rm e}^{-\bar q^{\rmH} - \bar q^\rmL}}{(h+1)^2} \cdot \frac{\ud h}{\ud \bar q^{\rmH}} \nonumber
\end{align}
 where $h$ is a shorthand notation for $h(\bar q^{\rmH} , \bar q^{\rmL})$. The first term of the above equation is positive for $q^{\rmH} \in (0,1)$. The sign of the second term is determined by $\frac{\ud h}{\ud \bar q^{\rmH}} $ which is nonnegative because $h$ can be rearranged as
$$
h(\bar q^{\rmH} , \bar q^{\rmL}) = \sum_{i=1}^{\Delta} {\rm e}^{\bar q^{\rmH} i} \cdot \left( 1 - {\rm e}^{- \bar q^{\rmH} - {\bar q^{\rmL}}}\right)
$$
which is nondecreasing in $\bar q^{\rmH}$. This completes the proof.

\newpage
\begin{IEEEbiographynophoto}{Jeong-woo Cho}
received his B.S., M.S., and Ph.D. degrees in Electrical Engineering and Computer Science from KAIST, Daejeon, South Korea, in 2000, 2002, and 2005, respectively. From September 2005 to July 2007, he was with the Telecommunication R\&D Center, Samsung Electronics, South Korea, as a Senior Engineer. From August 2007 to August 2010, he held postdoc positions in the School of Computer and Communication Sciences, \'Ecole Polytechnique F\'ed\'erale de Lausanne (EPFL), Switzerland, and at the Centre for Quantifiable Quality of Service in Communication Systems, Norwegian University of Science and Technology (NTNU), Trondheim, Norway. He is now an assistant professor in the School of Information and Communication Technology at KTH Royal Institute of Technology, Stockholm, Sweden. His current research interests include performance evaluation in various networks such as peer-to-peer network, wireless local area network, and delay-tolerant network.
\end{IEEEbiographynophoto}

\begin{IEEEbiographynophoto}{Jean-Yves Le Boudec}
is full professor at EPFL and fellow of the IEEE. He graduated from \'Ecole Normale Sup\'erieure de Saint-Cloud, Paris, where
he obtained the Agregation in Mathematics in 1980 (rank 4) and received his doctorate in 1984 from the University of Rennes, France. From 1984 to 1987 he was with INSA/IRISA, Rennes. In 1987 he joined Bell Northern Research, Ottawa, Canada, as a member of scientific staff in the Network and Product Traffic Design Department. In 1988, he joined the IBM Zurich Research Laboratory where he was manager of the Customer Premises Network Department. In 1994 he joined EPFL as associate professor.

His interests are in the performance and architecture of communication systems. In 1984, he developed analytical models of multiprocessor, multiple bus computers. In 1990 he invented the concept called ``MAC emulation'' which later became the ATM forum LAN emulation project, and developed the first ATM control point based on OSPF. He also launched public domain software for the interworking of ATM and TCP/IP under Linux. He proposed in 1998 the first solution to the failure propagation that arises from common infrastructures in the Internet. He contributed to network calculus, a recent set of developments that forms a foundation to many traffic control concepts in the Internet.

He earned the Infocom 2005 Best Paper award, with Milan Vojnovic, for elucidating the perfect simulation and stationarity of mobility models, the 2008 IEEE Communications Society William R. Bennett Prize in the Field of Communications Networking, with Bozidar Radunovic, for the analysis of max-min fairness and the 2009 ACM Sigmetrics Best Paper Award, with Augustin Chaintreau and Nikodin Ristanovic, for the mean field analysis of the age of information in gossiping protocols.

He is or has been on the program committee or editorial board of many conferences and journals, including Sigcomm, Sigmetrics, Infocom, Performance Evaluation and ACM/IEEE {\sc Transactions on Networking}. He co-authored the book {\it Network Calculus} (2001) with Patrick Thiran and is the author of the book {\it Performance Evaluation of Computer and Communication Systems} (2010).
\end{IEEEbiographynophoto}

\begin{IEEEbiographynophoto}{Yuming Jiang} received his BSc from Peking University, China, in 1988, MEng from Beijing Institute of Technology, China, in 1991, and PhD from National University of Singapore, Singapore, in 2001. He worked with Motorola from 1996 to 1997. From 2001 to 2003, he was a Member of Technical Staff and Research Scientist with the Institute for Infocomm Research, Singapore. From 2003 to 2004, he was an Adjunct Assistant Professor with the Electrical and Computer Engineering Department, National University of Singapore. From 2004 to 2005, he was with the Centre for Quantifiable Quality of Service in Communication Systems (Q2S), Norwegian University of Science and Technology (NTNU), Norway, supported in part by the Fellowship Programme of European Research Consortium for Informatics and Mathematics (ERCIM). Since 2005, he has been with the Department of Telematics, NTNU, as a Professor. He visited Northwestern University, USA from 2009 to 2010.

He was Co-Chair of IEEE Globecom2005 - General Conference Symposium, TPC Co-Chair of 67th IEEE Vehicular Technology Conference (VTC) 2008, and General/TPC Co-Chair of International Symposium on Wireless Communication Systems (ISWCS) 2007-2010. He is first author of the book ``{\em Stochastic Network Calculus}''. His research interests are the provision, analysis and management of quality of service guarantees in communication networks. In the area of network calculus, his focus has been on developing models and investigating their basic properties for stochastic network calculus (snetcal), and recently also on applying snetcal to performance analysis of wireless networks.
\end{IEEEbiographynophoto}

\end{document}